\documentclass[12pt,preprint]{aastex}

\lefthead{Brogan et al.}
\righthead{OH (1720 MHz) Masers Toward W51B/C}

\begin{document}
\newcommand\HCOp{HCO$^+$}
\newcommand\HII{H\,{\sc ii}}
\newcommand\HI{H\,{\sc i}}
\newcommand\tco{$^{13}$CO(2--1)}
\newcommand\Ca{C110$\alpha$}
\newcommand\kms{km~s$^{-1}$}
\newcommand\cmt{cm$^{-2}$}
\newcommand\cc{cm$^{-3}$}
\newcommand\Blos{$B_{los}$}
\newcommand\Bth{$B_{\theta}$}
\newcommand\Bm{$B_{m}$}
\newcommand\Bv{$\mid\vec{B}\mid$}
\newcommand\mum{$\mu$m}
\newcommand\muG{$\mu$G}
\newcommand\mjb{mJy~beam$^{-1}$}
\newcommand\jb{Jy~beam$^{-1}$}
\newcommand\dv{$\Delta v_{FWHM}$}
\newcommand\va{$v_A$}
\newcommand\Np{$N_p$}
\newcommand\np{$n_p$}
\newcommand\km{km~s$^{-1}$}
\newcommand\h{^{\rm h}}
\newcommand\m{^{\rm m}}
\newcommand\s{^{\rm s}}

\title{OH (1720 MHz) Masers: A Multiwavelength Study of the
  Interaction between the W51C Supernova Remnant and the W51B Star
  Forming Region}
 
\author{ C. L. Brogan\altaffilmark{1}, W. M. Goss\altaffilmark{2},
  T. R. Hunter\altaffilmark{1}, A. M. S. Richards\altaffilmark{3}, 
  C. J. Chandler\altaffilmark{2}, J. S. Lazendic\altaffilmark{4},
  B.-C. Koo\altaffilmark{5},
  I. M. Hoffman\altaffilmark{6}, \& M. J. Claussen\altaffilmark{2}}

\altaffiltext{1}{National Radio Astronomy Observatory, 520 Edgemont Rd, 
Charlottesville, VA 22903, USA; cbrogan@nrao.edu.}

\altaffiltext{2}{National Radio Astronomy Observatory, P. O. Box 0, Socorro, 
NM 87801, USA}
 
\altaffiltext{3}{Jodrell Bank Centre for Astrophysics, Turing
  Building, University of Manchester, Manchester M13 9PL, UK}

\altaffiltext{4}{Monash Unversity, Clayton, VIC 3800, Australia}

\altaffiltext{5}{Astronomy Program, SEES, Seoul National University,
  Seoul 151-742, South Korea}

\altaffiltext{6}{Wittenberg University, Springfield, OH 45501, USA}
 
\begin{abstract}

We present a comprehensive view of the W51B \HII\/ region complex and
  the W51C supernova remnant using new radio observations from the
  VLA, VLBA, MERLIN, JCMT, and CSO along with archival data from {\em
  Spitzer}, {\em ROSAT}, {\em ASCA}, and {\em Chandra}.  Our VLA data
  include the first $\lambda=400$~cm (74~MHz) continuum image of W51
  at high resolution ($88\arcsec$).  The 400~cm image shows
  non-thermal emission surrounding the G49.2-0.3 \HII\/ region, and a
  compact source of non-thermal emission (W51B\_NT) coincident with
  the previously-identified OH (1720~MHz) maser spots, non-thermal 21
  and 90~cm emission, and a hard X-ray source.  W51B\_NT falls within
  the region of high likelihood for the position of TeV $\gamma$-ray
  emission.  Using the VLBA three OH (1720 MHz) maser spots are
  detected in the vicinity of W51B\_NT with sizes of 60 to 300~AU and
  Zeeman effect magnetic field strengths of 1.5 to 2.2~mG.  The
  multiwavelength data demonstrate that the northern end of the W51B
  \HII\/ region complex has been partly enveloped by the advancing
  W51C SNR and this interaction explains the presence of W51B\_NT\/
  and the OH masers.  This interaction also appears in the thermal
  molecular gas which partially encircles W51B\_NT and exhibits narrow
  pre-shock (${\Delta}v\sim5$~\kms) and broad post-shock
  (${\Delta}v\sim20$~\kms) velocity components. RADEX radiative
  transfer modeling of these two components yield physical conditions
  consistent with the passage of a non-dissociative C-type
  shock. Confirmation of the W51B/W51C interaction provides additional
  evidence in favor of this region being one of the best candidates
  for hadronic particle acceleration known thus far.

\end{abstract}

 
\section{INTRODUCTION}

Star formation triggered by the interaction of supernova remnants
(SNRs) with their parent molecular clouds has long been thought to
play an important role in the production of new generations of
stars. Indeed, it is thought that the enrichment of our own Solar
System with heavy elements might have resulted from such an
interaction. In practice such interactions are very difficult to
observe due to the extreme kinematic complexity present toward the
inner Galactic plane where most massive stars and SNRs are found. Over
the last two decades it has been recognized that when OH (1720 MHz)
masers are found toward SNRs they act as signposts for SNR/molecular
cloud interactions \citep[see for
  example][]{Wardle2002,Frail1996,Green1997,Frail1998,Frail2011}.  To
date, OH (1720 MHz) masers have been found in $\sim$ 24 SNRs, or 10\%
of the known SNRs in our Galaxy
\citep{Frail1996,YZ1996,Green1997,Koralesky1998,Sjouwerman2008,Hewitt2009a}.
Indeed, all 24 OH (1720 MHz) maser SNRs can be found in the
\citet{Jiang2010} catalogue of 34 SNRs thought to be interacting with
associated molecular clouds.

The observational properties of these SNR masers are quite different
from those of \HII\/ region OH (1720 MHz) masers. SNR OH (1720 MHz)
masers have: (1) larger maser spot sizes; (2) narrow and simple line
profiles; (3) low levels of circular polarization (typically $<10\%$);
(4) low magnetic field strengths ($< 3$ mG); (5) relatively non-variable flux
densities; (6) a notable lack of any other maser species; and (7) low
luminosity
\citep{YZ1996,Claussen1997,Claussen1999,H2O,Claussen2002,Koralesky1998,Brogan2000,Hoff2003,W44,W28,Woodall2007,McDonnell2008,Lazendic2010,Pihlstrom2011}.
These observational facts can be explained if OH (1720 MHz) SNR masers
originate in the post-shock molecular gas behind C-type shocks, and
are collisionally pumped in contrast to their radiatively pumped
\HII\/ region counterparts \citep{Elitzur1976}.

Current theories further suggest that the collisional pump is most
efficient for post-shock densities of $\sim 1\times 10^5$ cm$^{-3}$
and temperatures in the range 50 K $\la T \la$ 125 K, and that the
abundance of OH must be enhanced by the dissociation of water due to
X-rays and/or $\gamma$-rays emanating from the SNR
\citep{Lockett1999,Wardle1999,Hewitt2008}.  Observations of the
physical conditions in the pre- and post-shock gas available to date
are in good agreement with these theoretical expectations \citep[see
  e.g.][]{Frail1998,Reach2005}. Prior to this work, however, there
have been few in-depth studies employing multi-transition,
multi-species analyses.  Interestingly, a number of recent
publications have also noted the growing number of $\gamma$-ray
detected SNRs that are interacting with nearby molecular
clouds. \citep[][to name a
  few]{Hewitt2009b,Castro2010,Mehault2011,Frail2011,Uchiyama2012}. Indeed,
many of the $\gamma$-ray SNRs are associated with well known OH (1720
MHz) sources: e.g. G349.7+0.2,CTB 37A, 3C391, W44, IC443 and
W51C. Hence, as a powerful probe of SNR/molecular cloud interactions,
observations of this OH maser line can potentially pinpoint sites of
particle acceleration.


\begin{figure*}
\plotone{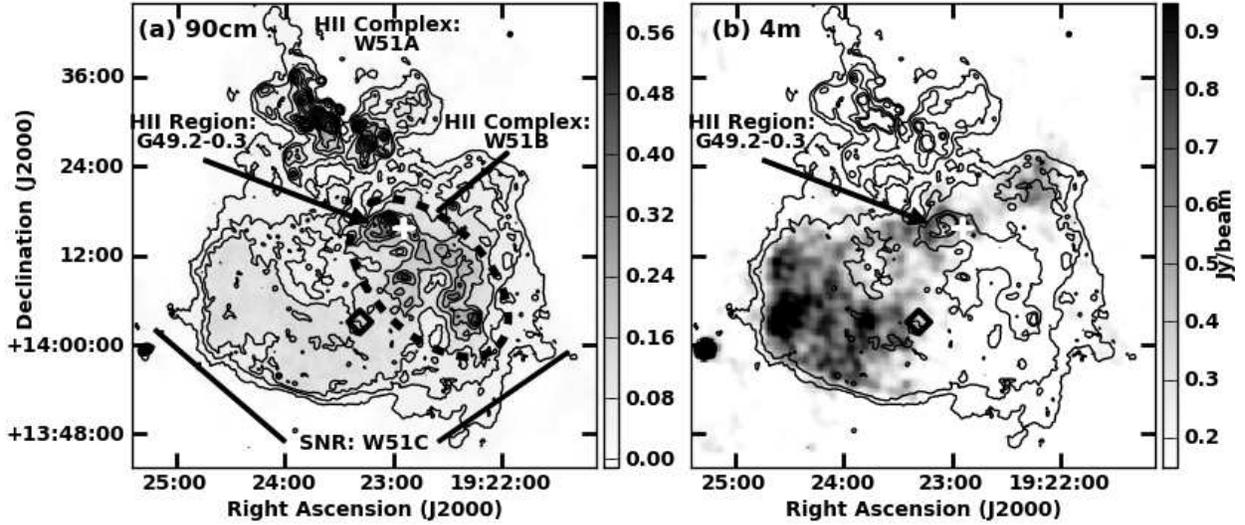} 
\caption[]{(a) VLA 90~cm B+C+D configuration image of the W51 complex
  (greyscale and contours). This image has a resolution of
  $35\arcsec\times 33\arcsec$ (PA=$-86\arcdeg$). The contour levels are
  30, 60, 90, 150, 210, 300, 400, \& 500 \mjb\/. Labels indicate the
  emission associated with W51A, W51B (dashed ellipse), and W51C.(b)
  VLA 400~cm B+C configuration image in greyscale with an angular
  resolution of $92\arcsec\times 84\arcsec$ (PA=$29.8\arcdeg$), with 90~cm contour
  levels of 30, 60, 150, 300, \& 400 \mjb\/.  On both panels, the
  white $+$ symbol shows the location of the OH (1720 MHz) masers, and
  the black diamond symbol shows the location of the PWN candidate CXO
  J192318.5+1403035 from \citet{KooChandra}. The W51B \HII\/ region
  G49.2-0.3 is also labeled.}
\label{fig1}
\end{figure*}

Of the known SNR OH (1720 MHz) masers, the origin of those observed
toward the W51 complex remain among the most poorly understood
\citep{Green1997,Brogan2000}.  W51 is composed of two large \HII\/
region complexes W51A (also known as W51 Main) and W51B, as well as
the SNR W51C (see the 90 and 400~cm images in
Figure~\ref{fig1}a,b). W51 is located near the Sagittarius arm tangent
point at $\sim l=49\arcdeg$ and $\sim b=0.3$ and thus velocity
crowding in this direction has made distance determinations (and
relative orientations along the line of sight) for the three W51
components (A, B, and C) uncertain.  In the past, their distances were
considered to be in the range of 5-7 kpc \citep{Kolpak2003}. A maser
parallax distance has recently been measured for W51A of 5.4 ($\pm
0.3$) kpc \citep{Sato2010}. Additionally, the soft X-ray absorption
seen toward the W51B region compared to the rest of the W51C SNR
suggests that W51B lies in front of W51C \citep{Koo1995}. However, it
remains uncertain whether W51A is in front of, co-distant with, or
behind W51B. The OH (1720 MHz) masers reported by \citet{Green1997}
are located $\sim 2.3\arcmin$ west of the W51B \HII\/ region
G49.2-0.3.  Subsequent observations of these masers by
\citet{Brogan2000} demonstrated that these masers exhibit all the
hallmarks of SNR OH (1720 MHz) masers.  Like Green et al., these
authors conclude that these masers are most likely excited by the
interaction between the W51C SNR (the only known SNR in the region)
and the molecular gas associated with the W51B \HII\/ region.

However, the exact nature of this interaction remains unclear
especially since SNR OH (1720 MHz) maser theory suggests that a
sufficient column of OH with the necessary velocity coherence can only
be achieved in shocks that move (more or less) perpendicular to the
line of sight \citep[see for example][]{Lockett1999}. It is difficult
to reconcile the need for an edge-on shock with the fact that W51B
lies in front of the W51C SNR (i.e. it seems that any shock would
propagate with a substantial component along the line of sight).  This
paper describes a multiwavelength investigation of the W51B/C OH (1720
MHz) masers and their environment spanning ten decades in wavelength
(400~cm to X-rays) and six decades of angular resolution (10~mas to
$1\arcdeg$).  The goal of this study is to better understand the
properties of the maser emission through high resolution
MERLIN\footnote{MERLIN is a UK national facility operated by
  University of Manchester on behalf of STFC.} and Very Long Baseline
Array\footnote{The National Radio Astronomy Observatory operates the
  Very Long Baseline Array and is a facility of the National Science
  Foundation operated under a cooperative agreement by Associated
  Universities, Inc.} (VLBA) observations, as well as to understand
the physical conditions that have led to the maser emission using Very
Large Array (VLA)\footnote{The National Radio Astronomy Observatory
  operates the Very Large Array and is a facility of the National
  Science Foundation operated under a cooperative agreement by
  Associated Universities, Inc.}  400~cm, 90~cm, and 20~cm data, James
Clerk Maxwell Telescope (JCMT)\footnote{The James Clerk Maxwell
  Telescope is operated by The Joint Astronomy Centre on behalf of the
  Science and Technology Facilities Council of the United Kingdom, the
  Netherlands Organisation for Scientific Research, and the National
  Research Council of Canada.} CO(3--2) and \tco\/ data, and Caltech
Submillimeter Observatory (CSO)\footnote{The Caltech Submillimeter
  Observatory is operated by the California Institute of
  Technology under cooperative agreement with the National Science
  Foundation (AST-0838261).} HCO$^+$(3--2) and HCN(3--2) data. For
comparison, we also make use of archival {\em Spitzer} mid-IR 8
\mum\/ data, as well as {\em Chandra}, {\em ROSAT}, and {\em ASCA}
X-ray data.

\section{OBSERVATIONS}

The observing parameters for all of the observations described in this
paper are listed in Table~\ref{observing}. Additional
details are provided below.

\subsection{MERLIN OH (1720 MHz) Maser Observations}

The OH (1720 MHz) masers were observed for approximately 18 hours (15
hrs on source) over the course of three days using the six antennas of
the MERLIN array (Lovell was not used). The Doppler tracking
corrections, in the LSR convention, were applied in the correlator. In
addition to right circular and left circular polarization data, the RL
and LR cross correlation data were also recorded. Absolute flux and bandpass
calibration were carried out using observations of 3C~84. Initial phase
calibration was obtained from periodic observations of
J1922+1530. Instrumental polarization leakage corrections were derived
from the unpolarized source 3C84, and 3C286 was used to set the
absolute polarization angle. The data were processed using the AIPS
software package.

After initial amplitude, delay, and phase calibration, several
iterations of self-calibration on the strongest (W51\_2) maser channel
were carried out and subsequently applied to the full spectral line
dataset.

\subsection{VLBA + Y1 OH (1720 MHz) Maser Observations}

The OH (1720 MHz) masers were observed for approximately 12 hours (6
hrs on source) over the course of two days using the ten antennas of
the VLBA plus one VLA (Y1) antenna. Both right circular and left
circular polarization data were recorded.  All calibration and imaging
for these data were carried out using the AIPS software. Absolute flux
calibration was determined from periodic system temperature
measurements coupled with existing antenna gain tables.  Bandpass and
delay calibration were performed using observations of the strong
continuum source J2253+1608 (3C~454.3). In order to obtain accurate
absolute positions and sensitivity to potentially weak maser emission,
these observations were phase referenced. Since no calibration source
suitable for VLBA observations was known at that time, a short VLBA
survey of near-by VLA calibrators was carried out prior to the target
observations. From this survey, the JVAS source J1911+1611
($19\h\/11\m\/58.2574\s\/$, $+16\arcdeg11\arcmin46.865\arcsec$ with an
uncertainty of $\sim 1$~mas) had the strongest compact emission and
was used to phase reference the maser data with a cycle time of 2
minutes (this source is now available in the VLBA calibrator
catalog). Corrections for Doppler tracking of the target data were
applied offline in AIPS.

After initial amplitude, delay, and phase calibration, several
iterations of self-calibration on the strongest (W51\_2a) maser
channel were carried out and subsequently applied to the full line
dataset. 

\subsection{VLA 21, 90, and 400~cm Continuum Observations}

We have obtained VLA observations from the B+C configurations at
400~cm and A+B+C+D configurations at 90~cm (see Table~1 for
details). The FWHP primary beam size of the 90~cm and 400~cm data are
$\sim 3\arcdeg$ and $\sim 11\arcdeg$, respectively. The data were
reduced and imaged using the low frequency wide-field imaging
techniques described in \citet[][and references
  therein]{Brogan2004}. Gain, bandpass, and absolute flux calibration
for the 90~cm data were carried out using J1924+334, Cygnus~A, and
3C~48 respectively. In order to best match the 400~cm data a 90~cm
image was constructed using only the B+C+D configurations and the
multiscale clean algorithm in AIPS, with scales of $\sim 40\arcsec$
(i.e. the intrinsic beam given the combination of data and weighting),
$120\arcsec$, and $480\arcsec$.  To achieve the highest angular
resolution, a 90~cm image was also constructed from the A+B+C+D
configuration data and multiscale clean with scales of $\sim
20\arcsec$ (i.e. the intrinsic beam given the combination of data and
weighting), $100\arcsec, 200\arcsec$, and $300\arcsec$.

\begin{figure*}
\plotone{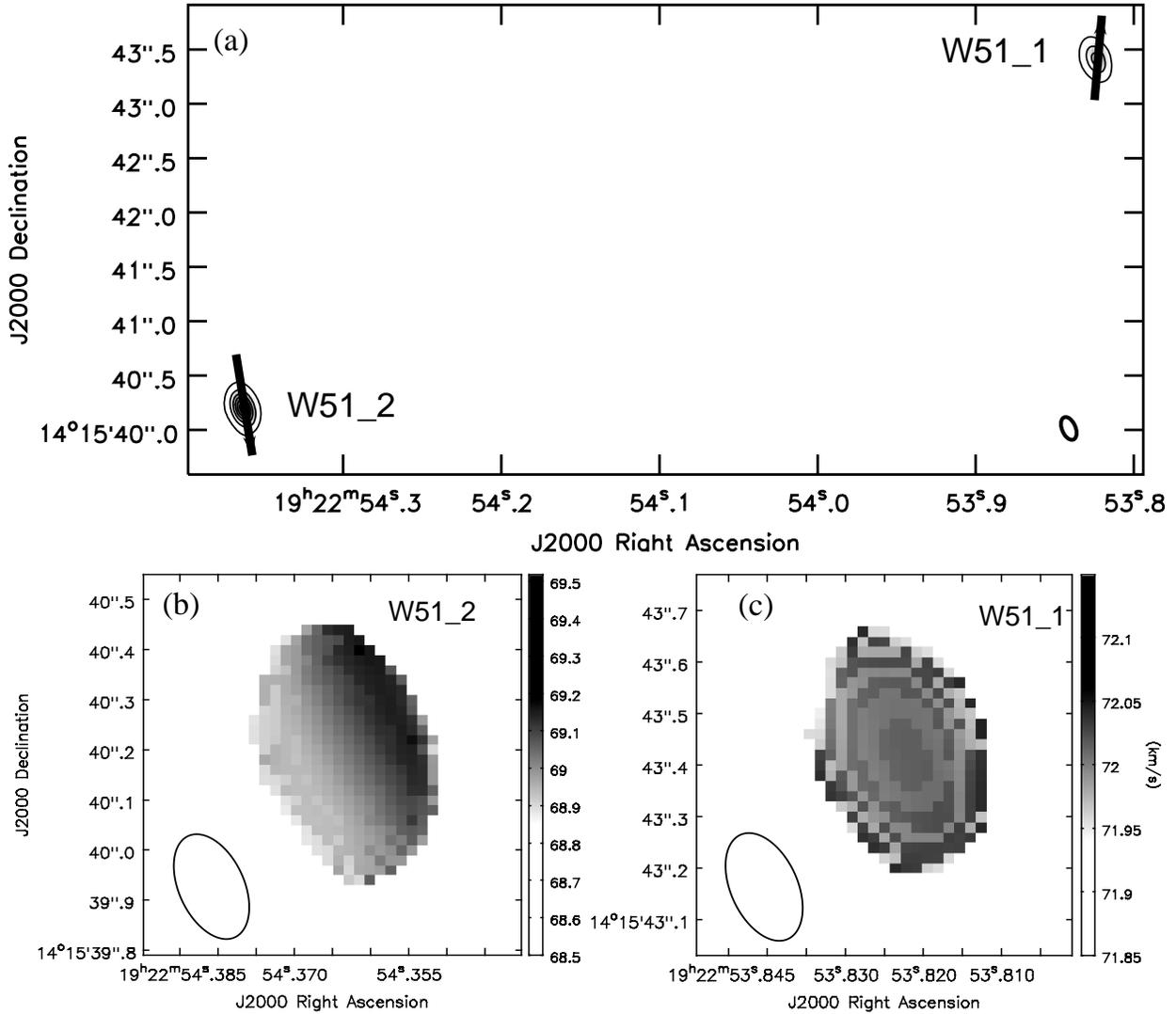}
\caption[]{(a) MERLIN integrated intensity image (greyscale and
contours) of maser spots W51\_1 and W51\_2. The contour levels are
100, 500, 1000, 1500, \& 2000 \mjb\/*\kms\/.  The beam with a size of
$221\times 125$ mas P.A.$=22.8\arcdeg$ is shown in the lower
right. The linear polarization position angle is also shown. (b) Zoomed
image of the 1st moment of W51\_2 (c) Zoomed image of the 1st moment of
W51\_1. Note the SE/NW velocity gradient of W51\_2, while W51\_1 shows
no velocity structure. Note that the velocity scales are different in 
(b) and (c).
\label{mermom}}
\end{figure*}

Absolute flux, bandpass, and coarse phase calibration at 400~cm were
all obtained from observations of Cygnus~A (located $\sim 28\arcdeg$
distant).  Of particular note is the lack of accurate absolute
position information for the 400~cm data due to the large distance to
the 400~cm phase calibrator. Typically this problem is solved by using
a 90~cm image as an initial model for the 400~cm
self-calibration. However, in the case of the W51 region, much of the
total 90~cm flux is dominated by \HII\/ regions making it unsuitable
for a 400~cm model (400~cm emission is absorbed by ionized thermal
emission along the line-of-sight creating distinct differences in
their morphology, see Fig.~\ref{fig1}a,b).  Therefore, the 400~cm
images have been shifted by $+20\arcsec$ in R.A. and $-10\arcsec$ in
Dec. so that the background point sources detected at both 400~cm and
90~cm (throughout the $5\arcdeg$ 90~cm primary beam) are coincident.

We have also obtained new VLA C-configuration 21~cm \HI\/ spectral
line data toward the W51B region (primary beam $\sim 30\arcmin$). The
continuum from the line free-channels of these data have been combined
with the continuum from line free-channels of archival VLA
D-configuration 21~cm \HI\/ data \citep{KooHI} to form a 21~cm
continuum image with a resolution matching that of the 90~cm B+C+D
configuration data. It should be noted, however, that owing to spatial
filtering the 21~cm data are missing flux on scales larger than $\sim
15\arcmin$ compared to $\sim 70\arcmin$ for the 90~cm data.

\subsection{JCMT CO (3--2) and $^{13}$CO (2--1) Observations}

In order to investigate the nature of the shocked molecular gas in the
vicinity of the OH (1720 MHz) masers we observed the CO(3--2) line
using the 345 GHz B3 receiver at the JCMT.  The
zenith opacity at 225~GHz measured by the tipping radiometer at the
CSO ranged from 0.10-0.12.  Data were taken in raster mode with 4-s
integrations per spectrum with samples separated by $5\arcsec$ in
right ascension and $7\arcsec$ in declination. The final raster is
$4\arcmin\times 5\arcmin$ in extent. Typical system temperatures were
between 400-500 K. The data were reduced using the SPECX software
package. A linear baseline was subtracted from each spectrum using
channels outside of the line emission region. In order to minimize
baseline ripples we used a nearby off-position that contained low
level CO(3--2) emission. To account for this emission we also observed
the reference position for 10 minutes in frequency switching mode, and
added the resulting emission back to each spectrum. In this paper, the
CO(3--2) line intensity is presented in units of $T_{MB}$ (K), where
$T_{MB}=T_a^*/0.63$. The final image cube was smoothed to $20\arcsec$
resolution.

A smaller region ($3\arcmin\times 2\arcmin$) around the OH (1720 MHz)
masers was also mapped in \tco\/ using receiver A3 when the zenith
opacity was 0.13. The samples were taken every $7\arcsec$ in right
ascension and every $10\arcsec$ in declination with 4-s
integrations. The system temperatures ranged from 250-300 K. The
angular resolution of these data is $20\arcsec$. For \tco\/, we used
$T_{MB}=T_a^*/0.69$.

\subsection{CSO HCO$^+$ (3--2) and HCN (3--2)Observations} 

We observed the maser position in HCN (3--2) at 265.886~GHz and
HCO$^+$ (3--2) at 267.558~GHz simultaneously using the Caltech
Submillimeter Observatory (CSO) in good conditions ($\tau_{\rm 225
  GHz} = 0.05$).  At these frequencies, the beam size is 26\arcsec\/
and the main beam efficiency is 0.7. Using position-switched mode, we
acquired 11~minutes of on source integration using the 230~GHz
receiver and the facility FFTS autocorrelators.  The reference
position was $19^{\rm h}22^{\rm m}44^{\rm s}.30,
+14\arcdeg\/05\arcmin\/50\farcs0$ (J2000) \citep{Ceccarelli2011}.  The
data were processed using the CLASS software package. The final
velocity resolution was smoothed to 0.60~\kms\/ for HCO$^+$ and
1.2~\kms\/ for HCN to improve the signal-to-noise. The results are
presented in $T_{MB}=T_a^*/0.70$.

\section{RESULTS}

\begin{figure*}
\plotone{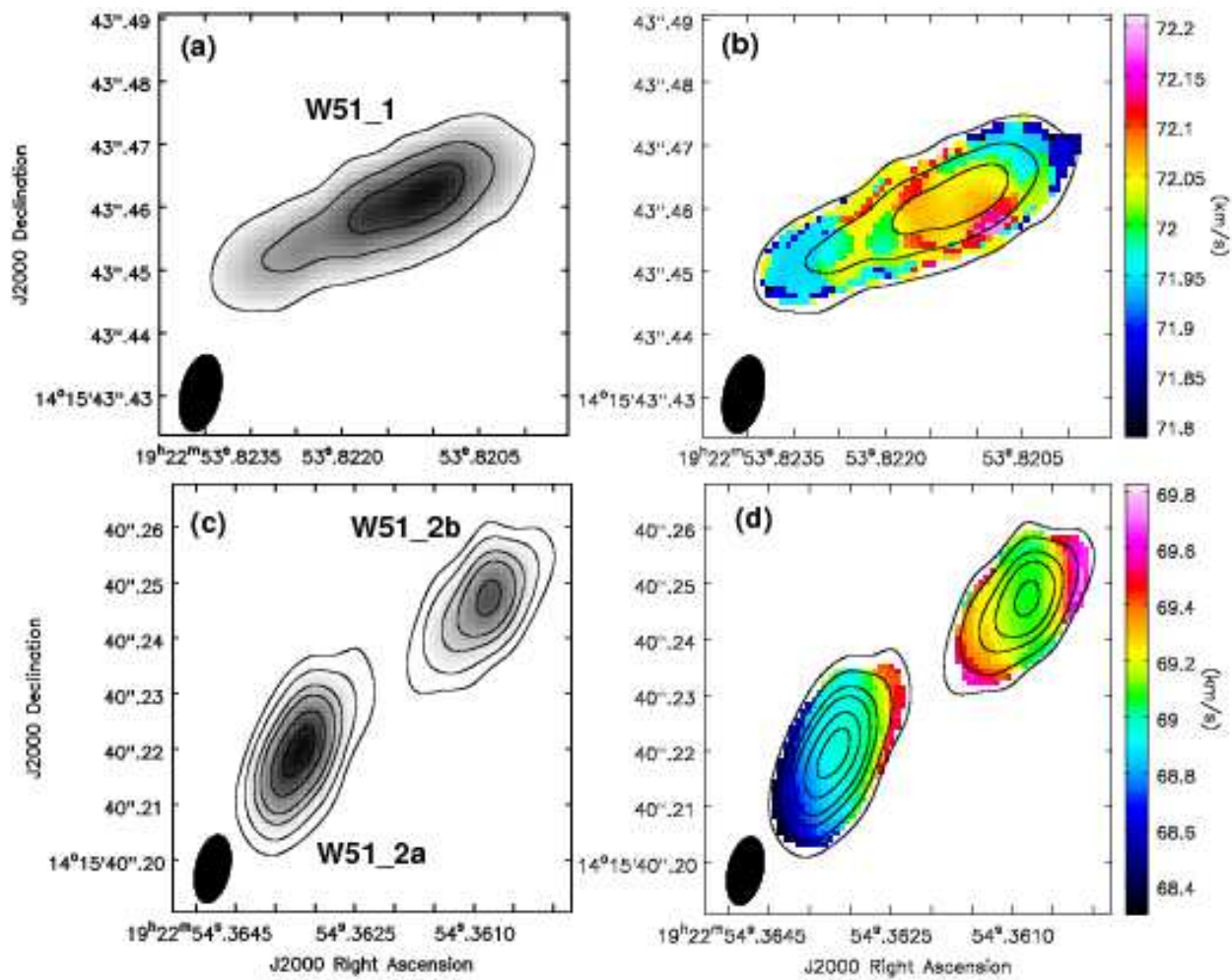}
\caption[]{ (a) VLBA integrated intensity image (greyscale and
  contours) of maser spot W51\_1. The contour levels are 50
  ($3\sigma$), 100, \& 150 \mjb\/*\kms\/. (b) Color image of the 1st
  moment (velocity gradients) superposed with the contours from
  (a). (c) VLBA integrated intensity image (greyscale and contours) of
  maser spot W51\_2. The contour levels are 50 ($3\sigma$), 100, 200,
  300, 500, 700, \& 900 \mjb\/*\kms\/. (d) Colorscale of the 1st
  moment (velocity gradients) superposed with the contours from
  (c). For all four plots the beam size of $12.5\times 6.3$ mas
  P.A.$=-5.2\arcdeg$ is shown in the lower left.
\label{vlbamom}}
\end{figure*}

\subsection{MERLIN and VLBA OH (1720 MHz) Data}

\subsubsection{Maser Spectral Line Properties}

Figure~\ref{mermom}a shows the MERLIN integrated intensity of the two
OH (1720 MHz) masers spots detected toward W51B. The locations of the
maser spots with respect to the W51 region is shown in
Fig.~\ref{fig1}.  The positions and central velocities are in good
agreement with those reported by
\citet{Frail1996,Green1997,Brogan2000} from VLA observations (see
Table~\ref{lines}). The results from fitting a Gaussian profile to the
maser spectra at the emission peaks are given in Table~\ref{lines},
along with the results from earlier VLA observations for reference
\citep{Brogan2000}.  At the resolution of the MERLIN data ($221\times
125$ mas, P.A.=$22.8\arcdeg$) the two maser spots remain
unresolved. However, maps of the velocity field of the two masers (see
Figs.~\ref{mermom}b, and c) are notably different, with the W51\_1
spot showing a moire pattern indicative of no resolved velocity
structure (note the very small width of the displayed velocities),
while the W51\_2 spot shows a clear SE-NW velocity gradient with
higher velocities to the NW.  The fitted MERLIN peak velocity and FWHM
line width is $72.018\pm 0.002$ \kms\/ and $0.902\pm 0.004$ \kms\/ for
W51\_1, while for W51\_2 the fitted values are $69.031\pm 0.001$ and
$1.283\pm 0.002$ \kms\/ (also see Table~\ref{lines}).  For comparison, at
assumed kinetic temperatures of 50 to 125 K, the thermal line width of
OH ranges from 0.3 to 0.6 \kms\/. Thus, roughly 1/2 to 1/3 of the
observed line width can be attributed to thermal motion, with the
remaining line width is due to bulk or non-thermal motions.

The integrated intensities and velocities (moment 1) of this region as
observed with the VLBA with a resolution of $12.5\times 6.3$ mas
($P.A.=-5\arcdeg$) are shown in Figures~\ref{vlbamom} a, b, c, and
d. These results are in excellent agreement with the MERLIN results --
W51\_1 remains a single, though elongated maser feature with little
velocity structure while W51\_2 is composed of two distinct maser
spots W51\_2a, and W51\_2b with velocities differing by only $\sim
0.09$ \kms\/ but with considerable velocity structure. The fitted line
parameters from the VLBA data are given in Table~\ref{lines}. The
fitted FWHM line width for W51\_1 is in excellent agreement with that
observed with MERLIN. The combined line width of the two components of
W51\_2 resolved by the VLBA is consistent with the MERLIN line width of
this maser, although the individual VLBA line widths of W51\_2a and
W51\_2b are slightly narrower.

We also fitted 2-D Gaussian components to the MERLIN and VLBA
integrated intensity images in order to measure the maser locations,
angular sizes, and flux densities. The results are given in Table~2.
The masers were unresolved by the VLA and so the peak intensity
$Sdv(peak)$ is equivalent to the integrated flux density $Sdv(int)$.
Interestingly, only about half of the VLA A-configuration integrated
flux density is recovered by the MERLIN observations, while the VLBA
and MERLIN recover a similar integrated flux density. If these masers
are non-time variable, this suggests that MERLIN and the VLBA resolved
out about half the total flux density. This suggests that these masers
have a ``core-halo'' morphology as seen previously for OH (1720 MHz),
CH$_3$OH, and H$_2$O masers \citep[see for example][]{Hewitt2008,
  Minier2002, Richards2011}. From the VLBA integrated intensity
images, the devonvolved fitted sizes of the maser spots are W51\_1:
$37.3\times 6.5$ mas (P.A.$112\arcdeg$), W51\_2a: $13.7\times 5.2$ mas
(P.A.$147\arcdeg$), and W51\_2b: $11.9\times 4.2$ mas
(P.A.$127\arcdeg$); with statistical uncertainties of about
$10\%$. For reference, at 6~kpc, 10~mas$\sim 10^{15}$ cm or $\sim 60$
AU. Using the peak intensities given in Table~\ref{lines} these sizes
correspond to peak brightness temperatures of $3.6\times 10^{8}$,
$6.0\times 10^{9}$, and $5.8\times 10^{9}$ K for W51\_1, W51\_2a, and
W51\_2b, respectively.


\subsubsection{Maser Polarization and Magnetic Field Properties}

From the MERLIN full Stokes polarization data, the maser spot W51\_1
has a polarized intensity ($p$=sqrt($Q^2+U^2)/I$) of $4.2\pm 0.4\%$ and a
position angle of $\chi^{maser}=-1\fdg6$, while the polarized
intensity of W51\_2 is $p=1.8\pm 0.3\%$ at $\chi^{maser}=4\fdg4$. The
uncertainty in the position angles ($\chi^{maser}$) are dominated by
the position angle calibration uncertainty which we estimate to be
about $5\arcdeg$. The MERLIN linear polarization position angles are
shown in Fig.~2a.  \citet{Brogan2000} reported a polarized intensity
of $3.5\%$ and $\chi^{maser}(\rm VLA)=-25\arcdeg$ for W51\_2, but the
polarization position angle calibration available for those data were
very uncertain.

Following \citet{Brogan2000}, we fit the Zeeman magnetic field
strength using the thermal Zeeman equation $V=0.5 Z$ \Bth $dI/d\nu$
where $V$ is Stokes $V(\nu)$, $Z$=0.6536 Hz \muG$^{-1}$, and $dI/d\nu$ is
the derivative of Stokes $I(\nu)$. The parameter \Bth$=C$\Bv, is the
magnitude of the magnetic field strength times a constant $C$ that may
depend on the angle $\theta$ between the magnetic field $\vec{B}$ and
the line-of-sight. For example, for thermal lines $C=cos(\theta)$; the
meaning of $C$ for the OH (1720 MHz) maser case will be discussed
further in \S4.3.

\begin{figure}
\plotone{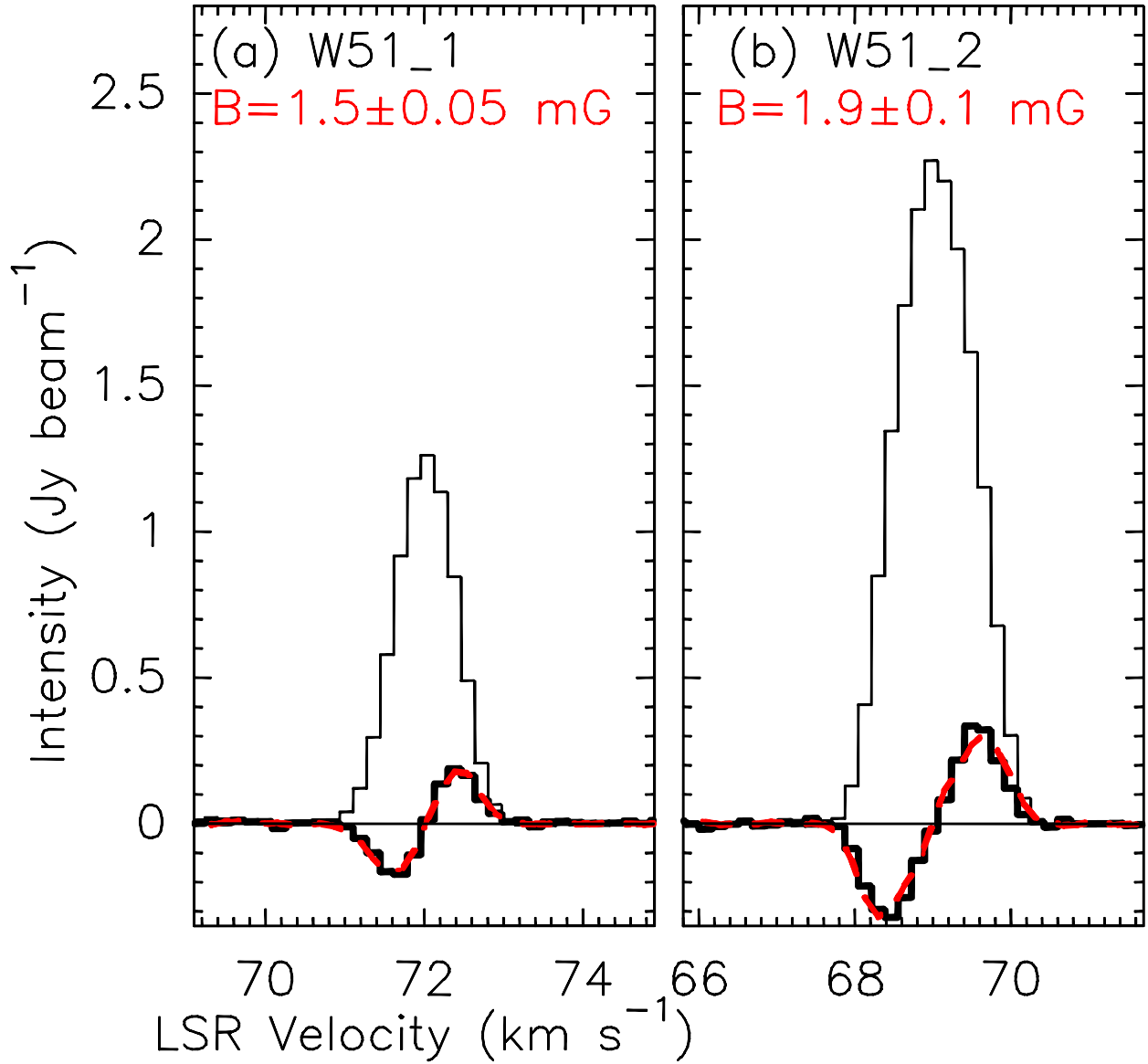}
\caption[]{Single pixel profiles of the MERLIN Stokes I (thin solid, top),
Stokes V (thick solid, bottom) and scaled derivative of Stokes I (dashed) at
the peak positions of the (a) W51\_1 and (b) W51\_2 masers. The fitted
magnetic field strengths and uncertainties are listed.
\label{B_mer}}
\end{figure}

\begin{figure}
\plotone{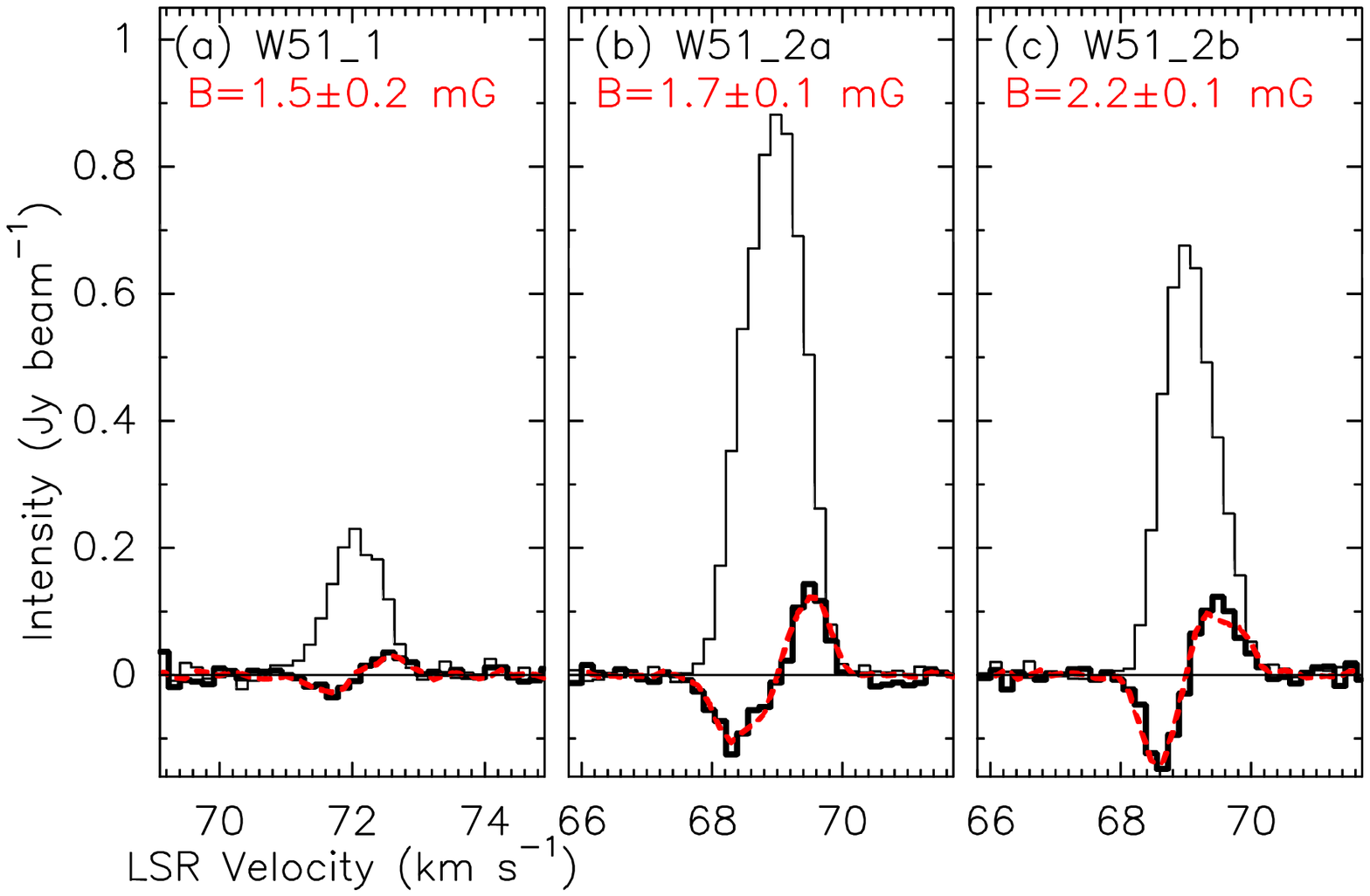}
\caption[]{Single pixel profiles of the VLBA Stokes I (thin solid, top),
Stokes V (thick solid, bottom) and scaled derivative of Stokes I (dashed) at
the peak positions of the (a) W51\_1, (b) W51\_2a, and (c)W51\_2b
masers. The fitted magnetic field strengths and uncertainties are
listed.
\label{B_vlba}}
\end{figure}

\begin{figure*}
\plotone{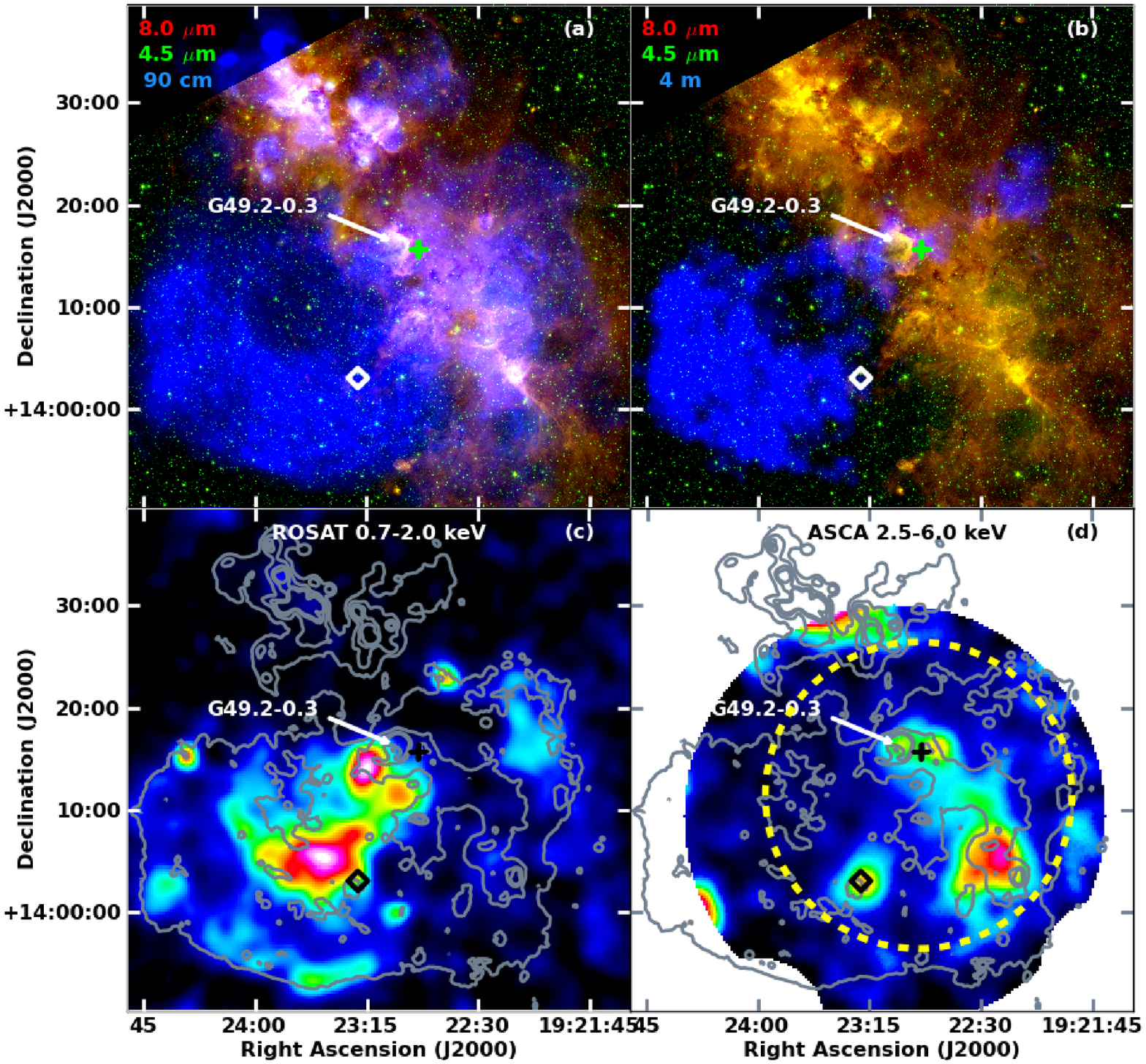}
\caption[]{ (a)-(b) Three color images of the W51 complex with red and
  green mapped to {\em Spitzer} 8.0 and 4.6 \mum\/ from the GLIMPSE
  survey, respectively. In (a), blue is mapped to VLA B+C+D
  configuration 90~cm data while in (b) blue is mapped to VLA B+C
  configuration 400~cm data. (c) {\em ROSAT} {\bf soft} X-ray emission
  from 0.7 to 2.5~keV (similar data are shown in \citet{KooASCA}). (d)
  {\em ASCA} {\bf hard} X-ray emission from 2.5 to 6.0~keV. The {\em
    ASCA} data are from \citet{KooASCA}. In (c) and (d), the grey
  contours are from the 90~cm image in (a) with levels of 60, 150,
  300, and 400 \mjb\/. The OH (1720 MHz) maser locations are marked by
  a single green or black $+$ symbol and the PWN candidate CXO
  J192318.5+1403035 from \citet{KooChandra} is marked by a black or
  white diamond. The location of the W51B \HII\/ region G49.2-0.3 is
  also indicated. Additionally, the region within which strong TeV
  emission was detected by MAGIC is indicated by the dashed yellow
  circle \citep{Aleksic2012}.
\label{big}}
\end{figure*}

Figures~\ref{B_mer} and \ref{B_vlba} show the Zeeman fits for the
MERLIN and VLBA data, respectively.  The fitted values of \Bth\ are
also listed in Table 2.  The signal-to-noise of these fits are
outstanding with S/N up to 30 (MERLIN for W51\_1) and 
S/N=7 (VLBA for W51\_1).  The observed values of \Bth\/ from 1.5 to
2.2~mG are in good agreement with the VLA observations of
\citet{Brogan2000}.  For the maser with the simplest velocity
structure, W51\_1, the MERLIN and VLBA magnetic field results are in
excellent agreement. For W51\_2, which the VLBA data reveal is
composed of two spots, the MERLIN value is essentially an intensity
weighted average of the two individual VLBA results.

These data are of sufficient quality that by following \citet{W44,W28}
we also independently derived \Bth\ by fitting Gaussian components to
the left and right circularly polarized profiles independently and
determining the line splitting directly,
i.e. \Bth=$(\nu_{RCP}-\nu_{LCP})/Z$. For example, for W51\_2a the line
splitting is $0.18\pm 0.01$ \kms\/. This method yielded consistent
results with the thermal Zeeman equation values listed in Table~2.

\subsection{VLA Radio Continuum}

The 90~cm (330 MHz) images presented in this work (for example
Fig~\ref{fig1}a) are qualitatively similar to the $2\arcmin$
resolution 90~cm image presented by \citet{Sub1995}, but the
resolution and sensitivity are significantly improved.
Figure~\ref{fig1}a shows the VLA B+C+D configuration 90~cm image of
the W51 region (resolution $\sim34\arcsec$, see Table~1) with the
three major components W51A (\HII\/ region complex), W51B (\HII\/
region complex), and W51C (SNR) identified. The OH (1720 MHz) masers
lie about $2\farcm3$ west of the W51B \HII\/ region G49.2$-$0.3, and
are coincident with an unresolved 90~cm source with an angular size of
$\sim 1\arcmin$ at a position of (J2000) $19{\rm h}22{\rm m}52.0{\rm
  s},+14\arcdeg15\arcmin50\arcsec$ (see Fig.~\ref{fig1}a).

\begin{figure*}
\plotone{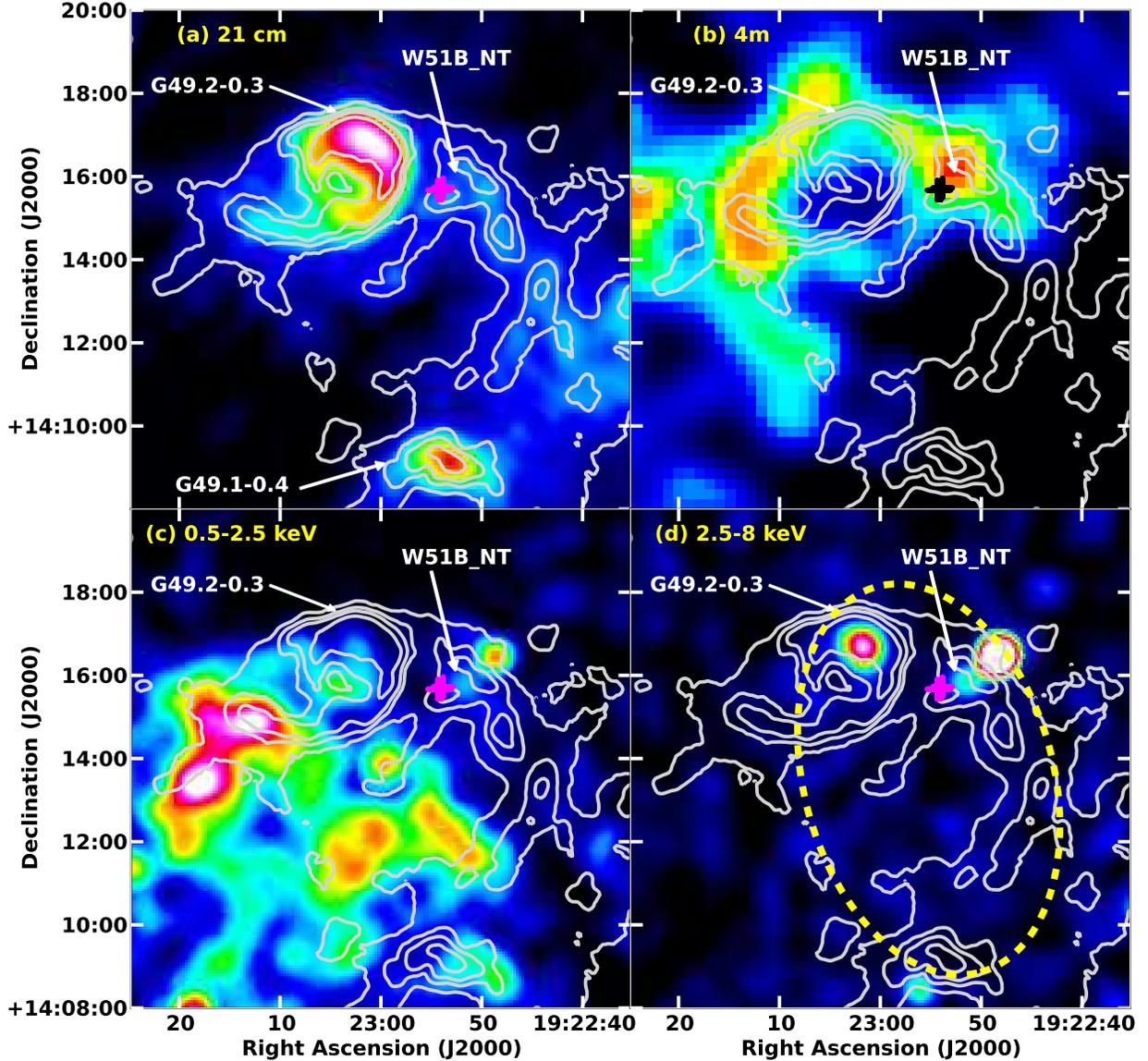}
\caption[]{(a) VLA 21~cm C+D configuration continuum image
  ($\sim34\arcsec$) using a log scale between -0.01 to 1.4 \jb\/. (b)
  VLA 400~cm B+C configuration image ($\sim90\arcsec$) using a linear
  scale between 0.15 and 1 \jb\/. (c) {\em Chandra} 0.5$-$2.5~keV
  image. (d) {\em Chandra} 2.5$-$8.0~keV image. For both (c) and (d) a
  linear scale between $1.7\times 10^{-8}$ to $1.4\times 10^{-7}$
  counts/s/pixel was used. All four panels are superposed with VLA
  90cm A+B+C+D configuration contours at 33, 55, 77, and 121 \mjb\/.
  The OH (1720 MHz) maser locations are marked with magenta or black
  $+$ symbols. The \HII\/ region G49.2$-$0.2 and the non-thermal radio
  source W51B\_NT are also labeled on each panel; the \HII\/ region
  G49.1$-$0.4 is also labeled on (a). On (d) we also show a ``test
  statistic'' contour level of approximately 6 from the MAGIC 100 -
  1000 GeV emission \citep[see Fig.~4 from][]{Aleksic2012}.}
\label{zoom}
\end{figure*}

The W51 region is shown for the first time at a wavelength longer than
2m in Figure~\ref{fig1}b. The VLA 400~cm B+C configuration image with a
resolution of $\sim 88\arcsec$ (see Table~1) reveals diffuse, extended
non-thermal emission concentrated mostly to the eastern side of the
W51C SNR, with some diffuse emission also appearing toward the
northwestern boundary of the W51C SNR. Most interesting for the
purpose of the current study is the non-thermal 400~cm emission that
appears to partially encircle the G49.2$-$0.2 \HII\/ region and the
discovery of an unresolved 400~cm source coincident with 90~cm emission
and the OH (1720 MHz) masers. Hereafter, we will call this source of
non-thermal emission W51B\_NT. It is notable that the area toward the
southern portion of the W51B string of \HII\/ regions is invisible at
400~cm, consistent with these regions being in the foreground of the W51C
SNR, and free-free absorbing the 400~cm emission from W51C along this
line-of-sight \citep[see for example][]{Brogan2005,Nord2006}.

To aid comparison of the thermal vs non-thermal gas,
Figures~\ref{big}a, b, shows three-color images constructed from the
Spitzer GLIMPSE 4.5 and 8~\mum\/ bands and the VLA 90~cm and 400~cm data,
respectively.  The mid-infrared emission traces thermal ionized gas,
as well as emission from dust and PAHs \citep{Benjamin2005}. Its
morphology matches closely that of the molecular cloud associated with
W51B \citep[see for example][]{KooCO,Bieging2010}. The 90~cm emission
(Fig.~\ref{big}a) traces both bright synchrotron (SNR) emission {\em
  and} optically thick free-free emission from \HII\/ regions. In
contrast, the 400~cm emission (Fig.~\ref{big}b) traces {\em only}
synchrotron emission except where it is absorbed by foreground
free-free emission. Fig.~\ref{big}c and d show the large scale {\em
  ROSAT} soft and {\em ASCA} hard X-ray emission, respectively, first
presented by \citet{KooASCA}.  Soft X-rays trace hot ($\sim 0.3$~keV)
thermal gas from the SNR, except where it is absorbed by the
intervening total column of atomic and molecular gas. In contrast, the
hard X-rays are mostly unaffected by the column of gas along the line
of sight and originates from even hotter thermal gas, including a
pulsar wind nebula candidate and young stars with fast winds.  We
detect diffuse 90~cm emission at the location of the pulsar wind
nebula candidate CXO J192318.5+1403035 from \citet[][the location is
indicated on Figs.~\ref{big}a-d]{KooChandra}, but it is not clear if
it arises from a distinct source compared to the overall diffuse SNR
emission. Similar to the 400~cm emission, there is a lack of strong soft
X-ray emission toward the southern portions of the W51B \HII\/ region
complex. In Fig.~\ref{big}d we also indicate the region in which
strong TeV emission was detected by the MAGIC telescope
\citep{Aleksic2012}.

\begin{figure*}
\plotone{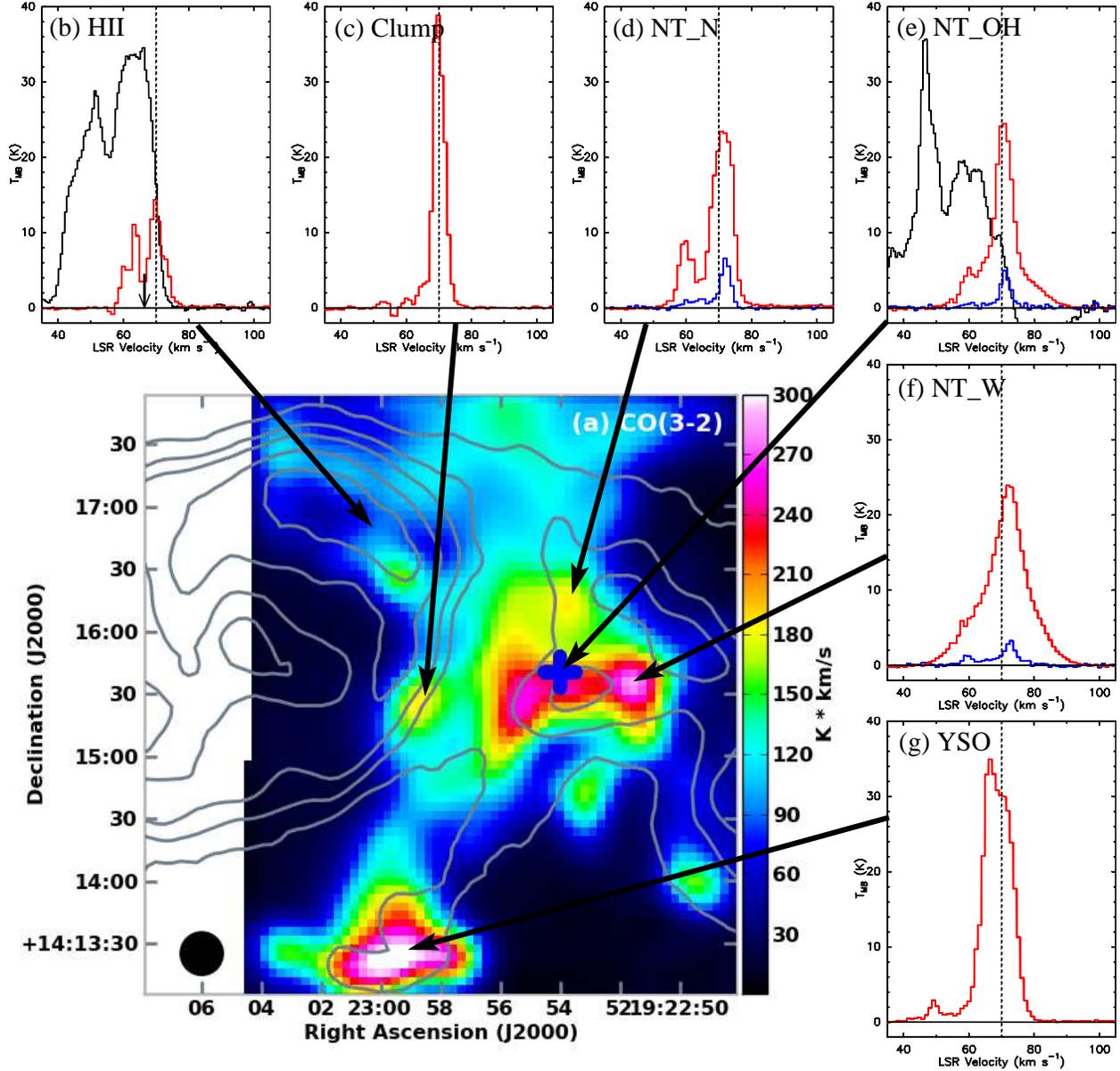}
\caption[]{(a) Color image shows the $20''$ resolution integrated
  intensity of the JCMT CO(3--2) emission (velocity range 47 to 97
  \kms\/) with black VLA A+B+C+D 90~cm continuum contours
  superposed. The 90~cm contour levels are 33, 55, 77, 121, and 165
  \mjb\/. The location of the OH (1720 MHz) masers is indicated by a
  blue $+$ symbol. The angular resolution of both images is
  $20\arcsec$ (beam is shown in lower left). (b)-(g) CO(3--2) (red),
  and where observed \tco\ (blue), spectral line profiles taken from
  the indicated J2000 positions.
Molecular emission appears to encircle W51B\_NT and
  a dramatic increase in the CO(3--2) line width is evident in this
  region. The arrow at 67.2 \kms\/ on the ``\HII\/'' spectrum
  indicates its radio recombination line velocity from
  \citet{Lockman1989}. C-configuration VLA \HI\/ absorption line
  profiles are also shown (black) on the ``\HII\/'' and ``NT\_OH''
  spectra (they have been multiplied by -100, and -1000,
  respectively).}
\label{CO}
\end{figure*}


A zoomed in view of the 21~cm continuum emission in the vicinity of
G42.9$-$0.3 and W51B\_NT is shown in Figure~\ref{zoom}a at
$\sim34\arcsec$ resolution. For comparison, the 400~cm emission and
archival soft and hard X-ray emission from {\em Chandra}
\citep[see][]{KooChandra} of the same field are shown in
Figures~\ref{zoom}b, c, and d. These images show the coincidence of
21, 90, and 400~cm emission toward W51B\_NT, as well as the presence
of a faint hard X-ray source at this position. Unfortunately, this
emission is too weak to extract a useful spectrum. The nature of the
stronger hard X-ray source to the NW of W51B\_NT is unknown.  After
accounting for the background contribution we estimate integrated flux
densities for W51B\_NT (defined by the 90~cm 0.15 \jb\/ contour shown
in Fig.~\ref{big}a, though it is partially obscured by the maser
symbol) of $\sim$ 0.5, 1.3, and 0.6 Jy at 400~cm, 90~cm, and 21~cm,
respectively. The uncertainty of these flux densities are dominated by
the background estimate, and are not accurate to better than $\pm 0.2$
Jy. The low 400~cm flux density is almost certainly a result of
free-free absorption along this complex line-of-sight \citep[see for
  example][]{Brogan2005}. The 21 and 90~cm flux densities imply a
non-thermal radio spectral index of $-0.7\pm 0.4$ ($S_{\nu}\propto
\nu^{\alpha}$).

Other groups have searched for possible non-thermal sources in the W51
complex. In particular, \citet{Moon1994} used the Bonn 11~cm Galactic
plane survey (resolution $4.2\arcmin$) along with IRAS 60 \mum\/
images to disentangle the thermal and non-thermal components of
W51. Though uncertain with respect to absolute flux densities, their
resulting images show the well-known thermal sources in W51A and W51B
as well as the non-thermal W51C SNR. Interestingly, they also see
evidence for a non-thermal source west of the G49.2-0.3 \HII\/ region.
More qualitative evidence for the existence of non-thermal emission in
the vicinity of W51B\_NT was presented by \citet{Copetti1991} at
151~MHz with a resolution of $2\arcmin \times 5\arcmin$ with a double
source detected near G49.2-0.3 with a separation of a few arcminutes
\citep[note that the published declination scale of the 151 MHz image
  must be displaced to the south by 5 arc min][]{Sub1995}. Though
suggestive, these early data have too poor an angular resolution to be
useful in the current analysis.


\subsection{Thermal Molecular Gas Properties}

The JCMT CO(3--2) integrated intensity from 40 to 95 \kms\/ with a
resolution of $20''$ is shown in Figure~\ref{CO}a.  Representative
spectra are shown in Figs.~\ref{CO}b-g), and where observed, the
$^{13}$CO(2--1) spectra are also shown (Figs.~\ref{CO}d-f). From these
data we find that CO emission partially encircles W51B\_NT with the
strongest emission along the eastern and southern boundaries. This
enhanced CO(3--2) emission is coincident with the locations of the OH
(1720 MHz) masers which are located towards the SE corner of the
W51B\_NT source (Fig.~\ref{CO}e, profile ``NT\_OH'' is taken from the
maser location). The spectra in the W51B\_NT region (Figs.~\ref{CO}d,
e, f) show at least three components: (I) a narrow ($\Delta v\sim 5$
\kms\/), but strong feature $T_{MB}\sim 20$ K at $\sim 71$ \kms\/;
(II) a broad ($\Delta v\sim 20$ \kms\/ wide) feature with a strength
of $T_{MB}\sim 10$ K also centered around 70 \kms\/; and (III) a
narrow ($\Delta v\sim 4$ \kms\/), weak blue-shifted feature at 60
\kms\/.  The appearance of the dramatically broadened component (II)
in the vicinity of W51B\_NT is strongly indicative of a shock.  The CO
throughout the mapped region has an average peak velocity of about
70~\kms\/ ($\pm 3$ \kms\/), in agreement with components (I) and (II)
toward the masers. Thus, while the broad velocity width observed
toward the W51B\_NT region is exceptional, it does not appear that the
W51B\_NT region arises from a distinct cloud along the line of site,
but is co-distant with the nearby \HII\/ regions like
G49.2$-$0.3. Additionally, the velocities of the narrow and broad CO
components at 70 and 71 \kms\/, respectively, are in excellent
agreement with the range of OH (1720 MHz) maser velocities (69-72
\kms\/; see Table~\ref{lines}).

Other features of note in the full CO(3--2) raster (see
Fig.~\ref{CO}b-g): (i) the emission associated with the G49.2$-$0.3
\HII\/ region is comparatively weak and is composed of two main
components that bracket the radio recombination line velocity from
\citet{Lockman1989} of 67.2 \kms\/. (ii) Along the SW rim of the
\HII\/ region is a compact ``clump'' with the strongest CO(3--2)
emission in the mapped region (40 K), but with narrow (4 \kms\/) line
widths. (iii) The southernmost CO(3--2) clump labelled ``YSO'' shows
evidence of bipolar outflow emission predominantly in an east-west
direction, but with a third component to the north. The ``YSO'' source
is also associated with weak 90~cm and 20 cm emission, though its
appearance is unremarkable in the mid-IR. From these clues, it seems
likely this source is an intermediate to massive protostar with the
radio continuum arising from optically thick free-free emission.

The appearance of the CO(3--2) and $^{13}$CO(2--1) spectra toward
W51B\_NT, with a broad ($\Delta v\sim 20$ \kms) component superposed
on a narrow ($\Delta v\sim 5$ \kms\/) feature at the LSR velocity is
in good agreement with other CO(3--2) studies of SNR OH (1720 MHz)
maser regions \citep[see for example][]{Frail1998,Reach2002}. To
facilitate comparison of all four observed thermal molecular line
species, we convolved the JCMT CO(3--2) and $^{13}$CO(2--1) cubes to
$26''$ resolution to match the CSO HCN(3--2) and \HCOp\/(3--2)
data. Spectra of all four species at this resolution toward the
centroid of the OH maser location are shown in
Figure~\ref{densitytemperature}a, and the parameters from Gaussian
fits using the CASA task specfit are given in Table~\ref{cofits}.

\section{DISCUSSION}

\subsection{Physical Conditions in the Pre- and Post-shock Gas}

\begin{figure*}
\resizebox{2.2in}{!}{\includegraphics{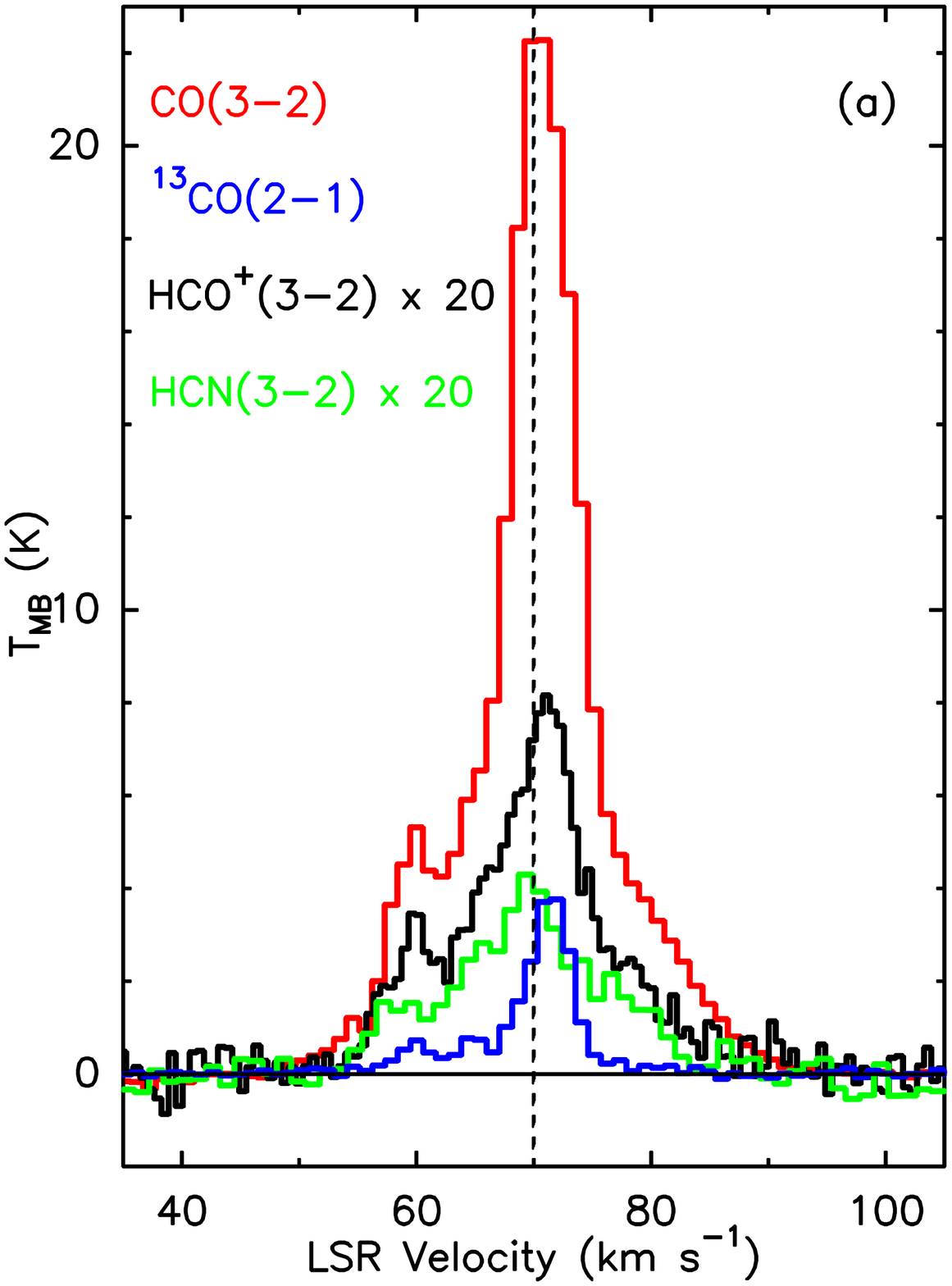}}
\resizebox{4.5in}{!}{\includegraphics{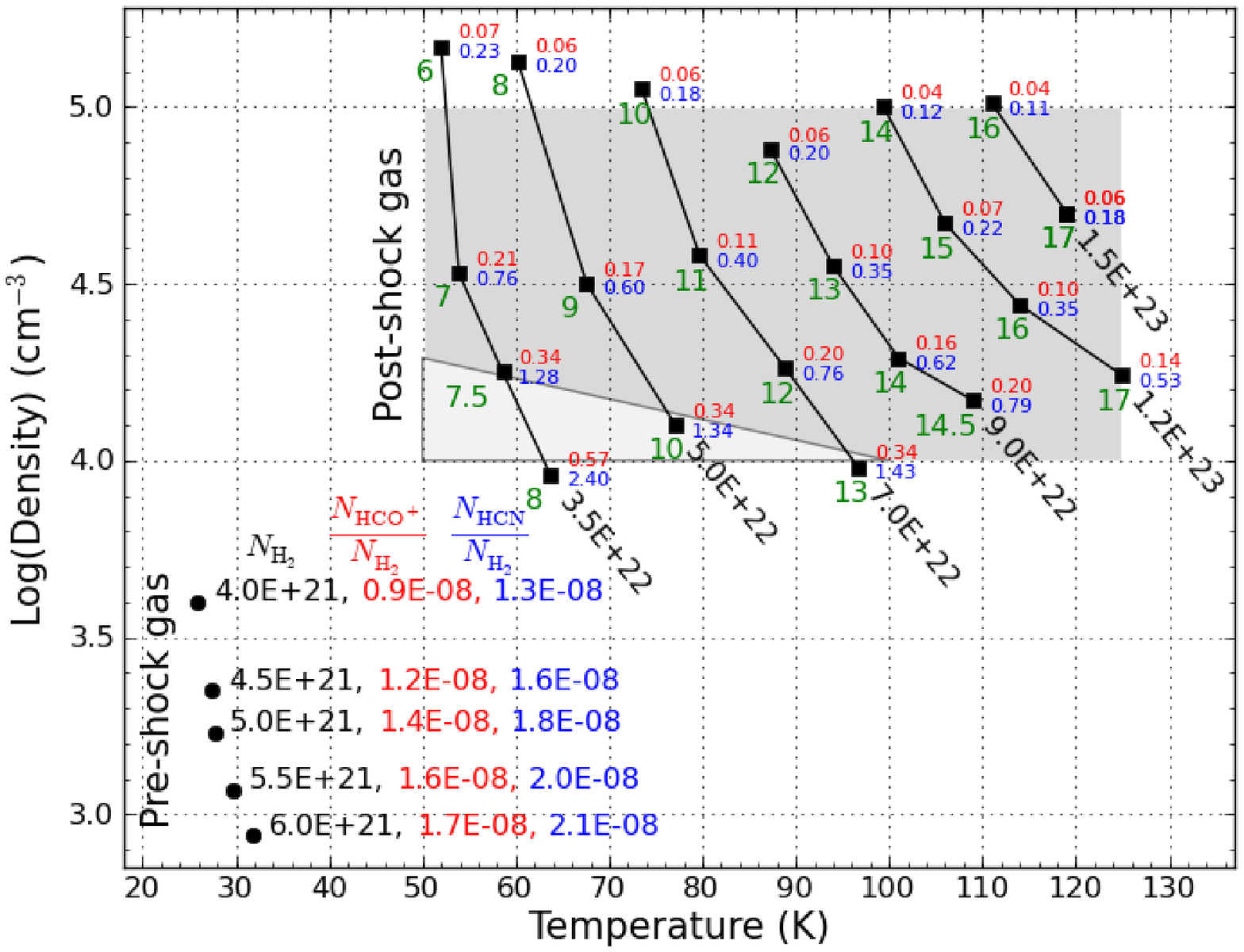}}
\caption[]{(a) Molecular line spectra with $26''$ resolution towards
  the maser position; the \HCOp\/ and HCN have been multiplied by 20
  to improve their visibility.  (b) Results of the RADEX radiative
  transfer models of the number density vs. temperature for the
  thermal gas toward the OH maser position based on the fitted CO and
  $^{13}$CO line profiles.  The gray shaded rectangular region
  indicates the theoretical range of parameters sufficient to pump the
  1720~MHz OH maser.  Filled circles denote the results for the narrow
  velocity component from the pre-shock gas and are labelled by the
  H$_2$ column density (assuming a beam filling factor of 1.0), and
  the corresponding abundances of \HCOp\/ and HCN.  Filled squares
  denote the results for the broad velocity component from the
  post-shock gas. The lines connect the results for the same H$_2$
  column density and the numbers in green to the left of the squares
  indicate the inverse of the beam filling factor. The numbers in blue
  and red denote the inferred HCN and \HCOp\/ abundances,
  respectively, in units of $10^{-8}$.}
\label{densitytemperature}
\end{figure*}

Using the $26''$ resolution fitted parameters at the OH maser position
(Table~\ref{cofits}), we ran the RADEX radiative transfer code
\citep{vandertak2007} with the LAMDA molecular data files
\citep{Schoier2005} to simultaneously model the CO(3--2) and
$^{13}$CO(2--1) line strengths and thereby estimate the physical
parameters of the pre-shock and post-shock gas.  Because these two
lines trace somewhat different excitation conditions, combining
observations of both lines helps to constrain the possible densities
and temperatures present.  We chose the slab option in RADEX as it is
the most appropriate geometry for both a shock and a molecular cloud
filament. We ran grids of RADEX models over density and temperature
for a range of plausible column densities, while assuming
N(CO)/N(H$_2$)$=10^{-4}$ and an isotopic abundance of
$^{12}$C/$^{13}$C=60 \citep{Milam2005}.  The H$_2$ column densities
and corresponding densities and kinetic temperatures that
simultaneously produce CO(3--2) and $^{13}$CO(2--1) antenna
temperatures that match the observed spectra are plotted in
Fig.\ref{densitytemperature}b.  Next we used RADEX to determine what
abundance of HCN(3--2) and \HCOp\/(3--2) would be required to produce
the observed intensities of these high density tracers at the
densities and kinetic temperatures derived from the CO observations.

Figure~\ref{densitytemperature}b summarizes the results for the four
species modeled. In the lower left corner of
Fig.~\ref{densitytemperature}b, the sequence of points of varying
H$_2$ column density (from $4 \times 10^{21}$ to $6 \times 10^{21}$
\cmt) indicate the conditions that reproduce the observed line
profiles in the pre-shock gas, with kinetic temperatures ranging from
26 to 32~K and H$_2$ densities of 850 to 4000~\cc.  The corresponding
\HCOp\/ and HCN line strengths are consistent with abundances of $1-2
\times\/ 10^{-8}$, which are typical of the values seen in massive
star formation regions \citep[][and references therein]{Doty02}.

We can compare the pre-shock column densities derived from the RADEX
analysis with those estimated from previous millimeter continuum and
X-ray observations toward this region. W51B\_NT is not detected at
1.1~mm in the $33''$ resolution Bolocam Galactic Plane Survey
\citep[the \HII\/ region G49.2-0.3 is detected; see][for Survey
  details]{Aguirre2011}. The $5\sigma$ upper limit measured in the
vicinity of W51B\_NT is about 0.36 \mjb\/ after multiplying by the 1.5
flux density correction factor recommended by \citet{Aguirre2011}.
Assuming the dust temperature is about $T_{\rm dust}=25$~K in the
pre-shock gas, a dust to gas ratio of 100, and a dust opacity of 0.0114
cm$^2$~g$^{-1}$ \citep{Ossenkopf1994,Aguirre2011}, we find a dust-derived $N({\rm
  H}_2)$ column density upper limit of $\sim 8\times 10^{21}$ \cmt. We
can also use the results of previous X-ray observations toward this
region to derive a lower limit. Using the {\em Suzaku} X-ray
satellite, \citet{Hanabata2012} find an average total proton column
density $N_p=N({\rm HI})+2N({\rm H}_2)$ toward their Region 3, which
encompasses most of W51B including W51B\_NT, of $2.4\times 10^{22}$
\cmt\/. Using a combination of Arecibo and VLA \HI\/ emission and
absorption data, \citet{KooHI} find $N(HI)\sim
1.9(T_{spin}/160~K)\times 10^{22}$ \cmt\/ along the line-of-sight to
the nearby \HII\/ region G49.2$-$0.3, with $80\%$ of the total column
originating close to the W51B region. Since $T_{spin}\sim 160$ K is an
upper limit, $1.9\times 10^{22}$ represents an upper limit to the true
N(HI).  When compared to the X-ray derived $N_p$, this suggests
$N({\rm H}_2)\geq 2.5\times 10^{21}$ \cmt\/. Thus, the allowed range
estimated from previous observations is $(2.5-8)\times 10^{21}$ \cmt\/,
compared to $N({\rm H}_2)=(4-6)\times 10^{21}$ \cmt\/ from the RADAX
molecular line analysis.

The post-shock gas is represented in the upper portion of
Figure~\ref{densitytemperature}b, with the gray shaded rectangular
region indicating the theoretical range of parameters sufficient to
pump the 1720~MHz OH maser.  A beam filling factor of significantly
less than 1 is required to explain the low observed line strengths in
the broad CO component.  This result is not surprising given that the
$26\arcsec$ beam corresponds to a linear diameter of $2.3 \times
10^{18}$~cm, because the width of the region of elevated temperature
in a typical interstellar C-type shock is $\sim 10^{16}-10^{17}$ cm,
depending on the ionization rate \citep{Wardle1999}.  The \HCOp\/
abundance is significantly lower (factors of 3-20) throughout the
shaded region than its value in the pre-shock gas, suggesting a
difference in chemistry.  Indeed, PDR models indicate that the
\HCOp/CO abundance ratio drops precipitously when the ionization rate
exceeds 10$^{-15}$~s$^{-1}$ \citep{Ceccarelli2011}.  As mentioned
previously, the W51B/W51C interface region is a strong emitter of
$\gamma$-rays with properties that are consistent with a hadronic
emission mechanism \citep[see][and references
  therein]{Aleksic2012}. \citet{Hewitt2009b} estimate $\gamma$-ray and
X-ray induced ionization rates of $4.4\times 10^{-16}$~s$^{-1}$ and
$8.8\times 10^{-16}$~s$^{-1}$, respectively, for this region. Though
dependent on many assumptions, together, the $\gamma$-ray plus X-ray
induced ionization rate is about 30 times the local interstellar rate,
and is in excellent accord with both that needed to produce an
adequate abundance of OH for the masers \citep{Wardle1999}, and that
required to lower the \HCOp/CO abundance as predicted by
\citet{Ceccarelli2011}.

Similar to \HCOp\/, the inferred abundance of HCN is also reduced in
much of the post-shock parameter space, with the exception of the
low-temperature, low-density end of the range (see
Fig.~\ref{densitytemperature}b).  While this suggests that the
physical conditions in the post-shock gas may lie in this triangular
region, it may be that the abundance of HCN is also altered by the
shock chemistry. In any case, a range of physical conditions almost
certainly exists across the shock, such that there is no single unique
answer. Still these observations demonstrate that the range of
expected densities and temperatures from the maser models are
supported by the thermal molecular line data.

\subsection{Nature of W51B\_NT}

\begin{figure*}
\plotone{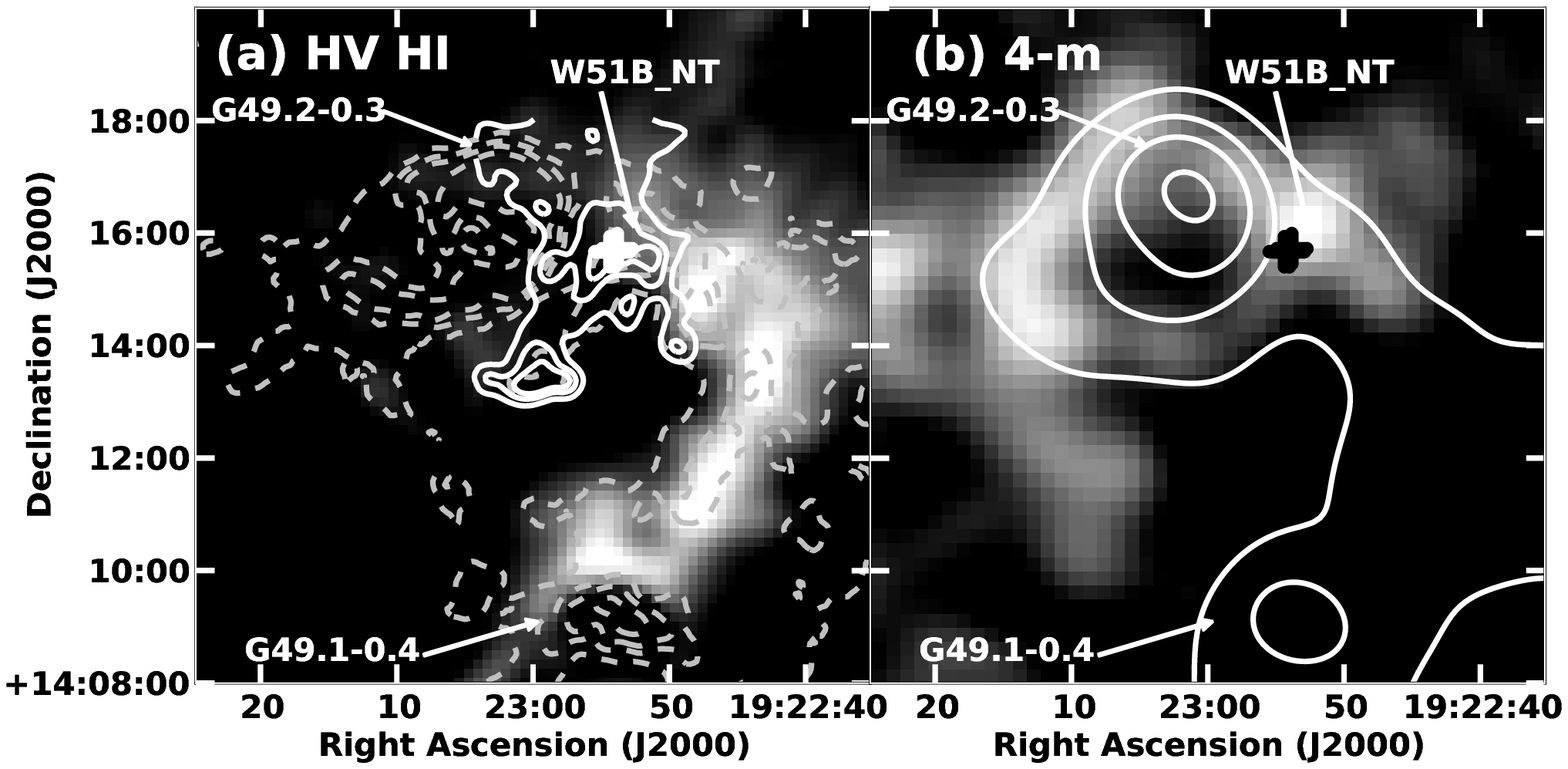}
\caption[]{(a) Greyscale \HI\/ integrated intensity image from 82 to
  140 \kms\/ showing the high velocity (HV) shocked \HI\/ emission
  from \citet{KooHI}.  Superposed are dashed VLA A+B+C+D 90~cm
  continuum contours at 11 $\times$ (3, 5, 7, 11, and 15) \mjb\/ and
  solid white JCMT CO(3--2) integrated intensity (from 47 to 97 \kms\/)
  contours at 60, 140, and 220 K*\kms\/ over its smaller field of
  view, see Fig.~\ref{CO}.  (b) Greyscale image of the 400~cm emission
  with a linear color scale from 0.3 ($3\sigma$) to 0.8 \jb\/, with
  white $\tau_{4m}$ contours at 1.5, 5, 10, and 20 superposed. The
  coincidence of detectable 400~cm emission where the predicted
  $\tau_{4m}$ is high suggests the \HII\/ region cannot be entirely in
  front of the W51C SNR. The locations of the OH (1720 MHz) masers are
  indicated by white (a) or black (b) $+$ symbols.}
\label{HVHI}
\end{figure*}

First, to summarize the key multi-wavelength results: a non-thermal
source of low frequency radio emission (W51B\_NT) has been detected in
the vicinity of the W51B OH (1720 MHz) masers. This source is
partially encircled by shocked molecular gas that is coincident in
space and velocity with the masers. Moreover, detailed modeling of the
molecular emission indicates that the physical conditions in the pre-
and post-shock gas are in good agreement with that expected from
collisional pump models. It is also notable that the molecular gas
associated with W51B\_NT shares a similar LSR velocity with nearby
\HII\/ regions (for example G49.2$-$0.3), suggesting that the masers
are co-distant with the northern members of the W51B \HII\/ region
complex. The non-thermal radio source is also coincident with a hard
X-ray source and falls within the region of high likelihood for the
position of TeV $\gamma$-ray emission, both of which have been
suggested as necessary to produce enough OH abundance (from water) in
C-type shocks for SNR OH (1720 MHz) maser emission. Thus, the evidence
indicates that these OH (1720 MHz) masers are of the SNR/molecular
cloud interaction variety.

The multi-wavelength data presented in this work suggests two possible
scenarios for the nature of W51B\_NT: (1) the non-thermal radio
continuum emission, CO shock, OH (1720 MHz) masers, and hard X-ray
emission are due to the W51C SNR, which lies close enough behind W51B
for its shock wave to have hit the \HII\/ region complex; or (2)
along this crowded, tangent point line of sight there is a previously
unknown small diameter ($\sim 2'$) supernova remnant co-distant
with the W51B \HII\/ regions. As will be described below, option (1)
seems the most consistent with the full range of data available for
this region.

It is clear from both 4m and soft X-ray absorption that the W51B
\HII\/ region complex lies at least partially {\em in front} of the
W51C SNR. The key uncertainty is whether W51B is close enough to W51C
for the SNR's shock front to have reached its back side. In addition
to the presence of W51B\_NT and the OH (1720 MHz) masers, there is
further evidence that the W51C SNR is interacting with much of the
northern end of W51B.  For example, \citet{KooHI} report a filamentary
arc of high velocity (HV), shocked \HI\/ emission at velocities
between $\sim 80-140$ \kms\/ toward the northern section of W51B.
This arc of emission begins just west of W51B\_NT, curves around to
the south, and ends just north of the \HII\/ region
G49.1$-$0.4. Figure~\ref{HVHI}a shows the morphology of the HV \HI\/
compared to the 90~cm continuum; the CO(3--2) emission from the
smaller field of view imaged with the JCMT is also shown. The arc of
the filament matches very closely with the morphology of the 90~cm
continuum emission in this region. \citet{KooHVCO} also find HV
CO(1--0) emission roughly co-spatial with the HV \HI\/ (most of the HV
CO(1--0) is outside the field of view of the CO(3--2) JCMT data, see
Fig.~\ref{HVHI}a).  These authors suggest that the HV gas is tracing a
shock interaction between the W51C SNR and the molecular gas
associated with the (foreground) W51B complex of \HII\/
regions. Indeed, the HV \HI\/ does appear to lie along the transition
between the unabsorbed 400~cm SNR emission and the large scale
filamentary molecular cloud associated with the W51B complex -- the CO
filament is roughly coincident with the {\em Spitzer} 8 \mum\/
emission shown in Fig.~\ref{big}a, b \citep[see for
  example][]{KooCO,Bieging2010}. \citet{KooHI} finds that the
properties of the HV \HI\/ shock are consistent with a dissociative
J-type shock, and further postulate that the post-shock gas is mostly
atomic since the HV \HI\/ column density is quite high. Therefore, the
non-detection of HV \HI\/ toward W51B\_NT is not surprising since it
is spatially coincident with the OH (1720 MHz) masers that are thought
to arise in non-dissociative C-type shocks \citep{Wardle1999}.

Another piece of evidence indicating a more extensive interaction is
the morphology of the 400~cm emission. In particular, the non-thermal
400~cm emission appears to both encircle and partially overlap with
the G49.2$-$0.3 \HII\/ region emission as defined by the shorter
wavelength 21~cm image (see Figs.\ref{zoom}a,b). As described, for
example, in \citet{Brogan2005} and \citet{Nord2006}, relatively little
thermal ionized gas is required to absorb 400~cm emission so it would
be surprising to detect such low frequency emission in the direction
of a foreground \HII\/ region. This expectation can be quantified by
comparing an estimate of the 400~cm free-free continuum opacity from
the \HII\/ region with the observed 400~cm emission. The 400~cm
opacity was calculated by first convolving the C+D 21~cm continuum
image to the 400~cm image resolution of $\sim 90''$, and then
converting this image to brightness temperature. Next the
$\tau_{21cm}$ was calculated from $T_b=T_e(1-{\rm exp}(-\tau))$
assuming an electron temperature of 10,000 K, and then extrapolated to
$\tau_{4m}$ using $\tau_{\lambda}=\tau_{21cm}(\lambda(cm)/21{\rm
  cm})^{2.1}$. Using this method, we find a peak 400~cm opacity of
22.5 with $\tau_{4m}>1$ extending well into the observed 400~cm
emission, see Figure~\ref{HVHI}b. Of course, in reality the $T_e$
within the \HII\/ region is unlikely to be constant, but the
relatively high assumed value yields a lower limit to $\tau_{21cm}$.
The strongest observed 400~cm emission is on the eastern side of W51C
at 1.35 \jb\/ (see Fig.~\ref{fig1}b). Since it seems unlikely that the
unabsorbed 400~cm emission underlying the \HII\/ region is
significantly brighter than this intensity, we take it to be an upper
limit to the unabsorbed 400~cm non-thermal emission toward
G49.2$-$0.3. When this assumption is coupled with the $3\sigma$ rms
noise of the 400~cm image of 0.3 \jb\/, we find that lines of sight
with $\tau_{4m}>1.5$ would be undetectable {\em if the \HII\/ region
  lies entirely in front of the 400~cm non-thermal emission}. However,
as demonstrated in Fig.~\ref{HVHI}b detectable 400~cm emission is
observed toward regions with $\tau_{4m}$ up to $\sim 20$!  This
analysis suggests, as does the morphology, that in fact the W51C SNR
has partially enveloped the G49.2$-$0.3 \HII\/ region.

Interestingly, from analysis of the optical extinction toward the W51B
region, \citet{Han2001} found that the northern half of W51B is more
heavily extincted than the south/southwestern end. Since the total
molecular column density is actually larger to the SW, these authors
\citep[also see][]{KooCO} suggest that the northern end of W51B is
tilted deeper into the cloud than the south/southwestern end, placing
it closer to W51C. From all of these comparisons it seems clear that
the W51C SNR is interacting with the northern part of the W51B \HII\/
region complex.

\subsection{Constraints on Magnetic Field Direction}

As a shock propagates, the gas is compressed in the direction the
shock is moving and the component of the magnetic field parallel to
the shock front will be amplified. For collisionally pumped maser
emission it is further expected that the shock must be moving more or
less perpendicular to our line-of-sight in order to assure adequate
velocity coherence, and column density along the line-of-sight. Thus,
collisionally pumped masers like OH (1720 MHz) masers associated with
SNRs and Class I CH$_3$OH masers are always found near the systemic
velocity. These two facts together imply that the magnetic field
vector is oriented somewhere in the plane perpendicular to the
direction the shock front is moving, and we are viewing that shock
front edge-on. Using this as the starting geometry, one would like to
use the polarization measurements to constrain the direction of the
magnetic field vector and the value of $C$ in the \Bth$=C$\Bv
equation.

Maser propagation theory \citep[see
  e.g.][]{Elitzur1998,Elitzur1996,Watson2001} suggests that the
observed percentage of linear polarization $p$ is a strong function of
the angle between the line-of-sight and the magnetic field vector
$\theta_{los}$. Indeed, for $\theta_{los} = 55\arcdeg$ the observed
$p$ should be zero (it is also zero for $\theta_{los} =
0\arcdeg$). Away from the critical angles of $55\arcdeg$ and
$0\arcdeg$, $p$ is non-zero, though its magnitude is a strong function
not only of $\theta_{los}$, but also the saturation level of the maser
and to some extent to the angular momentum of the molecular state that
is masing. The brightness temperatures of OH (1720 MHz) masers
$T_b\sim 10^{8-10}$ K, along with the lack of significant variability
are consistent with at least a moderate degree of saturation
\citep[also see additional arguments in][]{Brogan2000,W44,W28}.
Assuming moderate saturation, according to the model of
\citet{Watson2001}, for a J=2--1 state, $p$ is expected to be within
about $-10\%$ to $+20\%$, with positive values corresponding to
$\theta_{los}> 55\arcdeg$ and negative values corresponding to
$\theta_{los}< 55\arcdeg$. If $\theta_{los}$ is known, this model can
also be used to constrain the $C$ parameter (which may depend on
$\theta$) in the Stokes V conversion to magnetic field strength.  For
small values of $p$ and moderate saturation, $V\approx (Z/2) B
dI/d\nu$ (Z=the Zeeman coefficient, and $B$ is the total field
strength) with no $\theta$ dependence, i.e. $C\approx 1$ and hence
$B\approx B_\theta$. Currently, no detailed analysis is available for
the J=3/2 OH (1720 MHz) transition, though it would likely be
qualitatively similar to the J=2--1 case (private communication,
William Watson 2005). This suggests that given the small measured $p$
values of $+4.2\%$ and $+1.9\%$ (i.e. Stokes U is consistent with zero
intensity and Stokes Q is positive) for W51\_1 and W51\_2,
$\theta_{los}\gtrsim 55\arcdeg$. Unfortunately, unless a maser
polarization model is created for the J=3/2 OH (1720 MHz) transition a
more quantitative analysis is not be possible.

A maser's linear polarization position angle $\chi^{maser}$ can be
{\em either} parallel or perpendicular to the magnetic field in the
plane-of-the-sky $B_{\perp}$. However, maser theories also predict
that for $\theta_{los}> 55\arcdeg$, $\chi^{maser}\perp~B_{\perp}$
\citep[see e.g.][]{Elitzur1998,Watson2001}. As mentioned above, we
expect that \Bv\/ lies parallel to the plane of the shock front
(perpendicular to its direction of motion), and in principle the
orientation of the shock front itself can be discerned from the
morphology of the shocked CO(3--2) emission. Although there is
significant confusion from quiescent gas at the LSR velocity of the
W51B molecular cloud ($\sim 70$ \kms\/), the CO(3--2) emission from
the wings of the broad component (see Fig.~\ref{CO}e, f) appears to be
oriented East-West. This is also apparent from the CO(3--2) integrated
intensity image (Fig.~\ref{CO}a) which is dominated by the broad
component. An East-West orientation for the molecular shock front is
approximately perpendicular to that found for the linear polarization
position angle $\chi^{maser}$ (see \S 3.2.1; Fig.~\ref{mermom}a).
This result is in good agreement with the expectation described above
based on $p$ that $\theta_{los}> 55\arcdeg$ and hence we have
$\chi^{maser}\perp~B_{\perp}$. The situation where $\chi^{maser}$ is
perpendicular to the shock front is also seen in W28 but the opposite
orientation is observed for W44 \citep{W44,W28}. The fact that some
sources fall into the $\chi^{maser}\| B_{\perp}$ case and some the
$\chi^{maser}\perp B_{\perp}$ breaks the uncomfortable coincidence
pointed out by \citet{Brogan2000} that since all SNR OH (1720 MHz)
masers have linear polarization percentages of $\sim 10\%$ they could
all have essentially the same $B\|$ which is extremely unlikely.

A final consideration is whether the observed $p$ represents the true
polarized intensity.  The observed polarized intensity could be a
lower limit to the true intensity, owing to the factors mentioned by
\citet{Brogan2000}, including including Faraday depolarization and
tangling of the magnetic field lines within the masing region. Those
authors found that the Faraday depolarization length is only about
three times larger than the estimated maser gain length of about
$2\times 10^{17}$ cm \citep{Lockett1999}. The similarity of the gain
and Faraday depolarization lengths suggests that Faraday
depolarization is possible, but the new data do not allow further
insight on this issue. Luckily even in the presence of Faraday
depolarization the net polarization P.A. is not expected to change
\citep{Elitzur1992}. We can now exclude the second possibility down to
sizescales of $\sim 10^{15}$ cm (i.e. 10 mas at 6 kpc) based on the
fact that across 4 orders of magnitude in beam area from VLA to VLBA
resolution the observed Zeeman line splitting did not increase (except
as consistent with spectral blending) as would be expected if the
magnetic field were tangled on small size scales.



\section{SUMMARY \& CONCLUSIONS}

We have presented a wide range of new data toward the W51C SNR and the
W51B \HII\/ region complex. Using MERLIN and the VLBA we have
spatially resolved the OH (1720 MHz) masers in this region and
explored their magnetic field properties, establishing that (1) the
magnetic field strengths range from 1.5 to 2.2 mG (using the thermal
Zeeman equation); (2) the field strengths do not increase with angular
resolution (except where features were spatially unresolved and
spectrally blended by previous, lower-resolution observations)
suggesting the field is relatively smooth on these scales; (3) the
maser spots sizes are relatively large compared to \HII\/ region
masers, with linear dimensions of $(1-4)\times 10^{15}$ cm or $\sim
60-240$ AU at 6~kpc and brightness temperatures of $(0.4-6.0)\times
10^9$ K; (4) the linear polarized intensities are a few percent and
the position angle of the linear polarization is nearly zero; and (5)
the linear polarization properties suggest that the angle between the
magnetic field vector and the line-of-sight is $\gtrsim 55\deg$, and
that the difference between the magnetic field strengths measured
using the thermal equation and the total field strength is likely to
be small. More quantitative analysis requires a detailed polarization
model for the J=3/2 case.

We have presented the most sensitive, highest angular resolution
long wavelength images of this region to date at 90~cm and 400~cm
observed with the VLA showing the non-thermal radio continuum emission
from the W51C SNR with unprecedented detail. These data reveal (1) the
presence of non-thermal radio continuum emission in the vicinity of
the OH (1720 MHz) masers, which we denote W51B\_NT; (2) that the W51B
\HII\/ region complex must lie in front of the W51C SNR in agreement
with previous soft X-ray observations; and (3) that the nearby \HII\/
region G49.2$-$0.3 ($\sim 2.3'$ east of the masers) has been at least
partially enveloped by the W51C SNR by comparing the expected versus
observed 400~cm absorption. Through comparison with previous X-ray data,
we also find that a source of hard X-rays is coincident with W51B\_NT,
though the signal is too faint to model the spectrum.

Using observations of CO(3--2), $^{13}$CO(2--1), HCO$^+$(3--2), and
HCN(3--2) in the vicinity around the OH (1720 MHz) masers, we have
discovered a ring of shocked gas partially encircling the non-thermal
emission of W51B\_NT, coincident spatially and kinematically with the
OH (1720 MHz) masers. Radiative transfer modeling of the physical
conditions in the narrow velocity, unshocked gas yields a column
density of $(4-6)\times 10^{21}$ \cmt\/, a density of $(1-4)\times
10^3$ \cc\/, and a temperature of 26-32~K.  The broad velocity,
shocked gas is significantly smaller than the beam leading to a range
of possible column densities ($0.3-1.5 \times 10^{23}$ \cmt) and hence
physical conditions, but the most likely density and temperatures are
$1-2 \times10^4$ \cc\/ and 50-100~K, consistent with the passage of a
C-shock.

\acknowledgments

CLB thanks the JCMT and the University of Hawaii for fellowship
support during the course of this research. IMH thanks NRAO for
pre-doctoral fellowship support during the course of this project.  We
would also like to thank Professors William (Bill) Watson  and
Moshe Eliztur for helpful discussions about maser polarization
properties. We are also grateful for the help provided by Darek
Lis and Simon Radford in obtaining the CSO spectra.  This research has
made use of NASA's Astrophysics Data System.  We acknowledge the use
of NASA's {\em SkyView} facility (\url{http://skyview.gsfc.nasa.gov})
located at NASA Goddard Space Flight Center.

\newpage

\begin{deluxetable}{ll}
\tabletypesize{\footnotesize}
\tablewidth{0pc}
\tablecaption{Observing Parameters}
\tablecolumns{2}
\tablehead{
\colhead{Parameter}  & \colhead{Value} }     
\startdata
\cutinhead{MERLIN OH (1720MHz) masers}
Date & 2002 Jan 14, 15, 18 (MMO1B06)\\
Bandwidth & 0.25 MHz \\
Spectral channels & 256 \\
Channel separation & 0.17 \kms\/ \\
Velocity resolution & 0.24 \kms\/ \\
Spectral line rms noise$^a$ & 5 \mjb\/ \\
Synthesized beam & 221 mas $\times$ 125 mas P.A.$=22.8\arcdeg$ \\
\cutinhead{VLBA OH (1720MHz) masers} 
Dates & 2000 Dec 02,03, \& 04 (BB129)\\
Bandwidth & 0.25 MHz \\
Spectral channels & 256 \\
Channel separation & 0.17 \kms\/ \\
Velocity resolution & 0.24 \kms\/ \\
Spectral line rms noise$^a$ & 9 \mjb\/ \\
Synthesized beam & 12.5 mas$\times 6.3$ mas P.A.$=-5.2\arcdeg$ \\ 
\cutinhead{VLA 400~cm continuum}
Dates B-array & 2002 Jun 06 \& 22 (AB1031)\\
Date C-array & 2006 Oct 24 (AB1219)\\
Bandwidth & 1.5 MHz \\
Spectral Channels & 64 \\
Continuum rms noise & 100 \mjb\/ \\
Synthesized beam (B+C) & $92\farcs2\times 83\farcs5$ P.A. $29.8\arcdeg$ \\
\cutinhead{VLA 90~cm continuum} 
Date A-array & 2003 Aug 23 (AB1089)\\
Dates B-array & 2002 Jun 06 \& 22(AB1031) \\
Dates C-array & 2002 Nov 02; 2002 Dec 13 \& 30(AB1031)\\
Date D-array & 2003 Feb 08 (AB1077) \\
Bandwidth & 3.0 MHz \\
Spectral Channels & 32 \\
Continuum rms noise & 12 \mjb\/ \\
Synthesized beam (B+C+D) & $34\farcs8\times 32\farcs9$ P.A.$=-86\arcdeg$\\
Synthesized beam (A+B+C+D) & $20\farcs7\times 19\farcs8$ P.A.$=44\arcdeg$\\
\cutinhead{VLA 21~cm continuum and \HI}
  Date C-array & 2006 Nov 13 (AB1219)\\
  Date D-array (used for continuum only) & 1992 Jul (AK301)\\
  Bandwidth (C) & 3.0 MHz \\
  Spectral Channels (C) & 32 \\
  Continuum rms noise (C+D) & 5 \mjb\/ \\
  Spectral Line rms noise$^a$ (C) & 3 \mjb\/ \\
  Synthesized beam (C+D continuum) & $34\farcs8\times 32\farcs9$ P.A.$=-86\arcdeg$\\
  Synthesized beam (C \HI\/ Line) & $14\farcs7\times 14\farcs5$ P.A.$=14\arcdeg$\\
\cutinhead{JCMT 345 GHz CO(3--2)} 
Dates & 2004 Mar 29; Apr 20-25; Jun 11 \& 12 (m04ah45a2)\\
Rest frequency & 345.79599 GHz \\
Observing mode & Single sideband \\
Velocity resolution & 1.08 \kms\/ \\
Primary Beam & $14\arcsec$ \\
Spectral line rms noise$^a$ & 0.1 K \\
\cutinhead{JCMT 230 GHz $^{13}$CO(2--1)} 
Dates & 2005 Sep 22 (m05bh49c)\\
Rest frequency & 220.39868 GHz \\
Observing mode & Double sideband \\
Velocity resolution & 1.1 \kms\/ \\
Primary Beam & $20\arcsec$ \\
Spectral line rms noise$^a$ & 0.15 K \\
\cutinhead{CSO HCO+(3--2) \& HCN(3--2)} 
Dates & 2012 Nov 08 \\
Rest frequencies & 265.886 \& 267.558 GHz \\
Observing mode & Double sideband \\
Velocity resolution & 0.6 \& 1.2 \kms\/ \\
Primary Beam & $26\arcsec$ \\
Spectral line rms noise$^a$ & 0.02 \& 0.014 K \\
\enddata
\tablenotetext{a} {The rms noise per channel measured in main beam
  temperature $T_{MB}$.}
\tablenotetext{b} {Project codes are shown after the observation 
date(s).}
\label{observing}
\end{deluxetable}


\begin{deluxetable}{lcccccccc}
\tabletypesize{\tiny}
\tablewidth{0pc}
\tablecaption{Maser Spectral Line Parameters}
\tablecolumns{9}
\tablehead{
\colhead{Maser Spot}  & \colhead{Telescope} & \multicolumn{2}{c}{J2000 Coordinates\tablenotemark{b}} & 
\colhead{$V_{LSR}(peak)$} & \colhead{$\Delta V_{FWHM}(peak)$} & \colhead{$\int Sdv(peak)$} & \colhead{$\int Sdv(int)$} & \colhead{$B$}\\
&  & \colhead{$\alpha$ ($^{\rm h}~~^{\rm m}~~^{\rm s}$)} & \colhead{$\delta$ ($^{\circ}~~{\arcmin}~~{\arcsec}$)} &
\colhead{(\kms\/)} & \colhead{(\kms\/)} & \colhead{(\jb\/*\kms\/)} & \colhead{(Jy*\kms\/)}  & \colhead{(mG)}
}     
\startdata
W51\_1  & VLA\tablenotemark{a} &  &  &  72.0 (0.02)     & 0.9 (0.05)     & 2.6 (0.05)  & 2.6 (0.05)    & $1.5\pm 0.05$  \\
        & MERLIN  &  &                            &  72.018 (0.002)  & 0.902 (0.004) & 1.234 (0.004) & 1.38 (0.02) & $1.5\pm 0.05$  \\
        & VLBA    & 19 22 53.8210 & +14 15 43.462 &  72.06 (0.01)    & 0.901 (0.03)  & 0.211 (0.007) & 1.05 (0.05) & $1.5\pm 0.2$   \\
\hline
W51\_2  & VLA\tablenotemark{a} &  &  &  69.1 (0.02)     & 1.2 (0.05)     &  6.1 (0.05) &  6.1 (0.05)   & $1.9\pm 0.1$  \\
        & MERLIN  &  &                            &  69.031 (0.001)  & 1.183 (0.002)  &  2.939 (0.004) & 3.12 (0.02) & $1.9\pm 0.05$ \\
        & VLBA\_a & 19 22 54.3632 & +14 15 40.219 &  68.965 (0.002)  & 1.064 (0.004)  &  1.029 (0.004) & 2.07 (0.02) & $1.7\pm 0.1$  \\
        & VLBA\_b & 19 22 54.3608 & +14 15 40.247 &  69.050 (0.004)  & 0.957 (0.009)  &  0.700 (0.005) & 1.20 (0.04) & $2.2\pm 0.1$  \\
\enddata
\tablenotetext{a} {Data from \citet{Brogan2000}.}
\tablenotetext{b} {Position Uncertainties are $\sim 1$~mas. Only the most accurate VLBA positions are given, the VLA and 
MERLIN positions agree to within their individual (larger) uncertainties.}
\label{lines}
\end{deluxetable}

\begin{deluxetable}{lcccc}
\tablewidth{0pc}
\tabletypesize{\small}
\tablecaption{Fitted Parameters of Thermal Molecular Lines Toward OH Masers\label{cofits}}
\tablecolumns{5}
\tablehead{        \colhead{Transition\tablenotemark{a}} & \colhead{Component} & \colhead{Peak $v_{\rm LSR}$} 
                   & \colhead{Peak T$_{\rm MB}$} & \colhead{FWHM}\\
                   &                     & \colhead{(\kms)}      
                   & \colhead{(K)}        & \colhead{(\kms)}
}
\startdata 
$^{12}$CO 3--2  & narrow & 70.65 $\pm$ 0.03 & 18.57 $\pm$ 0.28 & 4.99 $\pm$ 0.09\\
               & broad & 70.12 $\pm$ 0.18 & 6.72 $\pm$ 0.26 & 18.64 $\pm$ 0.52\\
               & blueshifted & 59.40 $\pm$ 0.15 & 2.30 $\pm$ 0.26 & 3.00 $\pm$ 0.43 \\
\hline
$^{13}$CO 2--1  & narrow & 71.05 $\pm$ 0.07 & 4.31 $\pm$ 0.22 & 3.20 $\pm$ 0.20 \\
               & broad & 70.1\tablenotemark{a}  & 0.49 $\pm$ 0.16 & 15.47 $\pm$ 4.43\\
               & blueshifted & 59.4\tablenotemark{b}          & 0.43 $\pm$ 0.18 & 3.0\tablenotemark{b}\\
\hline
HCO+ 3--2      & narrow & 71.03 $\pm$ 0.16 & 0.20 $\pm$ 0.02 & 4.4 $\pm$ 0.5 \\
               & broad & 69.6 $\pm$ 0.3 & 0.21 $\pm$ 0.02 & 18.0 $\pm$ 1.0\\
               & blueshifted & 59.3 $\pm$ 0.3 &  0.07 $\pm$ 0.01 &3.0\tablenotemark{b}\\ 
\hline
HCN 3--2       & narrow & 69.6 $\pm$ 0.2 & 0.07 $\pm$ 0.01 & 1.7 $\pm$ 0.4 \\
               & broad & 69.7 $\pm$ 0.3 & 0.16 $\pm$ 0.02 & 14.9 $\pm$ 1.0\\
               & blueshifted & 57.4 $\pm$ 0.4 &  0.06 $\pm$ 0.01 &3.0\tablenotemark{b} 
\enddata
\tablenotetext{a} {Prior to fitting, the $^{12}$CO and $^{13}$CO spectral cubes were
  convolved to a beamsize of $26''$ in order to match the CSO HCO+
  spectrum. These fitted values were used in the modeling described in
\S4.1. }
\tablenotetext{b} {Values without uncertainties were fixed during the fit.} 
\end{deluxetable}

\end{document}